\documentclass[prd,floatfix,showpacs]{revtex4-1}
\usepackage{amsmath}
\usepackage{amsfonts}
\usepackage{amssymb}
\usepackage{graphicx}
\usepackage[margin=3.0cm]{geometry}
\usepackage{hyperref}
\usepackage{ dsfont }

\newcommand{\psih}{\hat{\psi}}
\newcommand{\no}[1]{\mathop{:}\nolimits\!\,#1\,\!\mathop{:}\nolimits}

\begin{document}
\title{Spectrum of the three dimensional fuzzy well}  
\date{\today}
\author{N Chandra$^1$, H W Groenewald$^2$, J N Kriel$^2$, F G Scholtz$^{2,3}$ and S Vaidya$^4$}
\affiliation{$^1$ The Institute of Mathematical Sciences, C.I.T. Campus, Taramani, Chennai - 600113, India}
\affiliation{$^2$ Institute of Theoretical Physics, University of Stellenbosch, Stellenbosch 7600, South Africa}
\affiliation{$^3$ National Institute for Theoretical Physics (NITheP), 7600 Stellenbosch, South Africa}
\affiliation{$^4$ Centre for High Energy Physics, Indian Institute of Science, Bangalore 560 012, India}

\begin{abstract}
We develop the formalism of quantum mechanics on three dimensional fuzzy space and solve the Schr{\"o}dinger equation for a free particle, finite and infinite fuzzy wells. 
We show that all results reduce to the appropriate commutative limits. 
A high energy cut-off is found for the free particle spectrum, which also results in the modification of the high energy dispersion relation.
An ultra-violet/infra-red duality is manifest in the free particle spectrum.
The finite well also has an upper bound on the possible energy eigenvalues.
The phase shifts due to scattering around the finite fuzzy potential well have been calculated.
\end{abstract}

\maketitle

\section{Introduction}
\label{intro}

Since the observation by Doplicher et al.\cite{Doplicher} that non-commutative space-time may be a scenario for space-time at short length scales that take into account quantum gravitational effects, the formulation of quantum mechanics \cite{scholtz1,Balachandran:2004rq} and quantum field theories \cite{doug} on such space-times has become a very active field of research.  The effects of non-commutative space-time are, however, only expected to show up at extremely high energies, temperatures and densities.  This motivates the study of the thermodynamic behaviour of non-commutative gases of fermions under these extreme conditions with the hope of establishing some benchmarks for possible experimental observation.  Using earlier results for the spectrum of a two-dimensional spherical well \cite{scholtz2}, the thermodynamics of Fermi gases confined by such a potential were investigated in \cite{scholtz3,scholtz4} as a first step.  This clearly showed sharp deviations from commutative behaviour at high density and pressure.  Most noteworthy of these results was the appearance of a high-energy cut-off in the spectrum, which resulted in 
incompressibility and non-extensive behaviour of thermodynamic quantities, such as entropy, at high densities \cite{scholtz3,scholtz4}.  

A natural question is how these observations manifest themselves in three dimensions. Here the most naive form of non-commutativity, i.e. $[x_i,x_j]=i\theta_{ij}$, proves to be problematic as it leads to a breaking of rotational symmetry due to emergence of a preferred commutative direction associated with the vanishing eigenvalue of the matrix $\theta_{ij}$. The preservation of the rotational symmetry was a key element for solving the spectrum of a non-commutative well in two-dimensions and the absence of rotational symmetry is a big stumbling block in generalising these results to three dimensions.  In addition, the thermodynamics of this scenario was already investigated in \cite{scholtz4} and found to lead to unwanted physical behaviour, which can in fact be taken as an argument against this simple form of non-commutativity. 

A way out of this dilemma is to start with a different model of non-commutativity in three dimensions that does not violate rotational symmetry.  The best known and simplest commutation relations that satisfy this condition, and that we adopt here, are the fuzzy sphere commutation relations defined by $[x_i,x_j]=i\epsilon_{ijk}x_k$.  For brevity we refer to this as the three dimensional (3D) fuzzy space. To pursue the program above in this setting, one has to define and solve for the spectrum of the spherical well in three dimensions. Some work in this direction was already undertaken in \cite{press} where the Coulomb problem in 3D fuzzy space was investigated.  

The aim of this paper is to formulate quantum mechanics on 3D fuzzy space, solve the free particle problem, define the notion of a non-commutative well (referred to as the 3D fuzzy well) and, finally, to solve for the spectrum of the well.  These results will be used in a forthcoming publication \cite{scholtz5} to compute the thermodynamics of a Fermi gas confined to such a well and explore the physical consequences.  

This paper is organised as follows: In section \ref{one} we briefly list some results for the free particle problem in three dimensional commutative space that are useful and of relevance to the non-commutative formulation pursued in section \ref{two}.    Section \ref{three} is devoted to solving the non-commutative free particle problem in 3D fuzzy space.  The resulting spectrum is analysed in section \ref{four}. Section \ref{five} discusses the problem of a spherical fuzzy potential well, which is defined, solved and analysed in this section. Section \ref{six} presents a useful complimentary treatment of the 3D fuzzy well based on the technique developed in \cite{press,Galikova:2011zf}, which is then used to solve the infinite well. Some useful results and their derivations are listed in the appendices.

\section{Commutative free particle}
\label{one}

Although a text book problem, it is useful to collect here, for future reference and comparison, some results for the commutative free particle problem.  In radial coordinates the Hamiltonian is given by
\begin{equation}
H_0 = \frac{1}{2m_0 r^2}\left(L^2+\hbar^2\square\right),
\end{equation}
where $L^2=\sum_iL^2_i$ and $L_i$ are the standard orbital angular momentum operators.   Furthermore $\square$ is a second order differential operator given in terms of the radial variable by
\begin{equation} \label{square}
\square = - \Delta (\Delta + 1), \quad \quad \Delta=r\frac{\partial}{\partial r}.
\end{equation}  
The simultaneous eigenfunctions of $L^2$ and $L_3$ are the well-known spherical harmonics $Y_{lm}(\theta,\phi)$ satisfying:
\begin{equation}
L^2 Y_{lm} = l(l+1) \hbar^2 Y_{lm}, \quad L_3 Y_{lm} = m\hbar Y_{lm},\quad \ell=0,1,2\ldots,\;m=-\ell,-\ell+1\ldots\ell.
\end{equation}
The free particle time independent Sch\"{o}dinger equation is
\begin{equation}
\label{eqn_sch_free}
H_0 \psi = E \psi.
\end{equation}
Setting
\begin{equation}
\psi= g_{k,l}(r) Y_{lm},
\end{equation}
yields the equation for the radial part of the wave function $g_{k,l}(r)$:
\begin{equation} \label{eqn_gr_comm}
r^2 g''_{k,l}(r) + 2rg'_{k,l}(r) + \left\{k^2 r^2 - l(l+1)\right\} g_{k,l}(r) = 0, \quad \quad \quad
k = \frac{1}{\hbar} \sqrt{2m_0 E}.
\end{equation}
One can write the solutions of the above equation in terms of Bessel polynomials \cite{krall}
\begin{equation} \label{radial_part_bessel_polynomial}
g_{k,l}(r) = \frac{e^{\pm ikr}}{kr} y_l\left(\pm \frac{i}{kr}\right).
\end{equation}
When expanded in powers of $x$ they read
\begin{equation} \label{bessel_expansion}
y_l(x) = \sum_{s=0}^{l} \frac{(l+s)!}{(l-s)!s!} \left(\frac{x}{2}\right)^s.
\end{equation}
These polynomials satisfy the differential equation
\begin{equation} \label{Bessel_polynomial_eqn}
x^2 y_l'' + 2(x+1)y_l' - l(l+1)y_l = 0.
\end{equation}
Note that each of the solutions given by (\ref{radial_part_bessel_polynomial}), i.e.,
\begin{eqnarray}
g_{k,l}^{(1)}(r) = \frac{e^{ikr}}{kr} y_l\left(\frac{i}{kr}\right), &\quad \quad \quad&
g_{k,l}^{(2)}(r) = \frac{e^{- ikr}}{kr} y_l\left(- \frac{i}{kr}\right),
\end{eqnarray}
is not well behaved at the origin.  However, one can construct a linear combination such that it is well behaved everywhere.
These are the spherical Bessel functions:
\begin{equation}
g_{k,l}(r) = j_l(kr) = \frac{i^{-l-1}}{2} g_{k,l}^{(1)}(r) - \frac{i^{l+1}}{2} g_{k,l}^{(2)}(r).
\end{equation}
In terms of these, the appropriate eigenfunctions of the full time-independent Schr\"odinger equation (\ref{eqn_sch_free}) reads
\begin{equation}
\psi_{klm}(r,\theta,\phi)=j_l(kr)Y_{lm}(\theta,\phi),
\end{equation}
with corresponding eigenvalue $E=\frac{\hbar^2k^2}{2m_0}$.  

In the non-commutative case it turns out that the mixed spherical harmonics $\mathcal{Y}_{lm}$ defined by 
\begin{equation}
\mathcal{Y}_{lm} = r^l Y_{lm}
\end{equation}
are the more natural quantities.  They satisfy
\begin{equation}
L^2 \mathcal{Y}_{lm} = l(l+1) \hbar^2 \mathcal{Y}_{lm}, \quad \quad L_3  \mathcal{Y}_{lm} = m \hbar \mathcal{Y}_{lm},
\end{equation}
as well as
\begin{equation}
\label{H0_mixedYlm}
\square \mathcal{Y}_{lm} = -l(l+1) \mathcal{Y}_{lm},\quad H_0 \mathcal{Y}_{lm} = 0.
\end{equation}
Writing the wave function as
\begin{equation}
\psi_{k,l,m} = h_{k,l}(r) \mathcal{Y}_{lm},
\end{equation}
or, in terms of $g_{k,l}(r)$,
\begin{equation} \label{gr_hr}
g_{k,l}(r) = r^l h_{k,l}(r)
\end{equation}
the equation for $h_{k,l}(r)$ is found to be
\begin{equation} \label{eqn_hr_comm}
r h''_{k,l}(r) + 2(l+1) h'_{k,l}(r) + k^2 r h_{k,l}(r) = 0.
\end{equation}
The two independent solutions are
\begin{eqnarray} 
h_{k,l}^{(1)}(r) = \frac{e^{ikr}}{kr^{l+1}} y_l\left(\frac{i}{kr}\right), &\quad \quad \quad&
h_{k,l}^{(2)}(r) = \frac{e^{- ikr}}{kr^{l+1}} y_l\left(- \frac{i}{kr}\right) \label{hr_bessel_polynomial}.
\end{eqnarray}
The well-behaved solution is a linear combination of the above two solutions and is given in terms of the spherical Bessel functions by
\begin{equation}
h_{k,l}(r) = \frac{j_l(kr)}{r^l} = \frac{i^{-l-1}}{2} h_{k,l}^{(1)}(r) - \frac{i^{l+1}}{2} h_{k,l}^{(2)}(r).
\end{equation}

A notion that appears quite naturally in the non-commutative three dimensional space based on fuzzy sphere commutations relations is that of the Hopf fibration, which is a map from $\mathds{C}^2 \sim \mathds{R}^4$ to $\mathds{R}^3$ given by
\begin{equation}
X_i = \frac{1}{2}\bar{z}_\alpha \sigma^i_{\alpha\beta} z_\beta.
\end{equation}
One can easily verify that
\begin{equation}
r^2 = X_i X_i = X_1^2 +X_2^2 + X_3^2 = \left(\frac{\rho^2}{2}\right)^2,
\end{equation}
$\rho$ being the radial distance in $\mathds{C}^2$
\begin{equation}
\rho^2 = \bar{z}_\alpha z_\alpha = |z_1|^2 + |z_2|^2.
\end{equation}
One can check that the operator $\frac{1}{2}\left(\bar{z}_\alpha\frac{\partial}{\partial \bar{z}_\alpha} + z_\alpha \frac{\partial}{\partial z_\alpha}\right)$ in $\mathds{C}^2$ corresponds to the radial ``scaling" operator $\Delta$ in $\mathds{R}^3$, i.e.,
\begin{equation}
\Delta = r \frac{\partial}{\partial r} = \frac{1}{2}\left(\bar{z}_\alpha\frac{\partial}{\partial \bar{z}_\alpha} + z_\alpha \frac{\partial}{\partial z_\alpha}\right).
\end{equation}

\section{Non-commutative formalism}
\label{two}

To set up the quantum formalism for a particle moving in 3D fuzzy space, we follow the approach of \cite{scholtz1}.  In this approach the first step is to identify an appropriate Hilbert space ${\cal H}_c$ (referred to as configuration space in \cite{scholtz1}), which carries a representation of the coordinate algebra.  In the next step the quantum Hilbert space ${\cal H}_q$, in which the pure states (`wave-functions') of the system are presented, is identified with the Hilbert-Schmidt operators on the configuration space that are generated by the coordinate operators.  The rationale behind this identification is the analogy with the standard formulation of quantum mechanics on the Hilbert space of square integrable wave-functions, i.e., we require the `wave-functions' to be functions of coordinates only and therefore to be generated by the coordinate operators, while the condition that the `wave-functions' are Hilbert-Schmidt is analogous to square integrability.  

The three dimensional fuzzy sphere is described by operators $\hat{X}_i$ satisfying the $SU(2)$ commutation relations
\begin{equation}
[\hat{X}_i,\hat{X}_j] = i\theta\epsilon_{ijk}\hat{X}_k,
\end{equation}
where $\theta$ is the non-commutative parameter with units of a length.   The Casimir $\hat{X}^2 = \hat{X}_i \hat{X}_i$ commutes with each $\hat{X}_i$ and is naturally identified with the radius of the fuzzy sphere and obviously  determined by the $SU(2)$ representation under consideration.  

It should now be clear how to identify the configuration space.  Since we wish to describe the fuzzy $\mathds{R}^3$, which can be thought of as a collection of spheres, the natural non-commutative configuration space is a Hilbert space that carries each allowed $SU(2)$ representation exactly once, i.e., we imagine non-commutative $\mathds{R}^3$ as an `onion' structure consisting out of a collection of fuzzy spheres:
\begin{equation}
{\cal H}_c={\rm span}\left\{|j,m\rangle,\;j=0,\frac{1}{2},1,\ldots,\;m=-j,-j+1,\ldots, j\right\}
\end{equation}
and the linear span is over the field of complex numbers.  

It is now easy to identify the quantum Hilbert space: the operators acting on ${\cal H}_c$ are elements of ${\cal H}_c\otimes {\cal H}^*_c$ (${\cal H}^*_c$ denotes the dual of ${\cal H}_c$) and in bra-ket notation a general linear combination of $|j^\prime,m^\prime\rangle\langle j,m|$.  However, to be an element of the quantum Hilbert space, these operators must be generated from coordinate operators only and, since these commute with the Casimir, the elements of the quantum Hilbert space must in addition commute with the Casimir, i.e., must be diagonal in $j$.  The quantum Hilbert space is therefore
\begin{equation}
{\cal H}_q=\left\{\psi=\sum_{j,m^\prime,m} c_{j,m^\prime,m}|j,m^\prime\rangle\langle jm|: {\rm tr_c}\left(\psi^\dagger\left(\hat{X}^2 + \frac{\theta^2}{4}\right)^{1/2}\psi\right)<\infty\right\},
\end{equation}
where ${\rm tr_c}$ denotes the trace over configuration space.  In bra-ket notation, we denote the elements of ${\cal H}_q$ by $|\psi)$ to distinguish them from elements of ${\cal H}_c$ and the inner product on ${\cal H}_q$ is given by 
\begin{equation}
\label{innprod_X}
(\psi|\phi) = 8\pi \theta^2 \, {\rm tr_c}\left(\psi^\dagger\left(\hat{X}^2 + \frac{\theta^2}{4}\right)^{1/2}\phi\right).
\end{equation}

The factor of $\left(\hat{X}^2 + \frac{\theta^2}{4}\right)^{1/2}$, which commutes with all the elements of $\mathcal{H}_q$ plays an important role in the hermiticity of the free particle Hamiltonian defined later in this section.
Observables are identified with self-adjoint operators acting on the quantum Hilbert space.  Obvious observables are the coordinates with actions defined by left multiplication:
\begin{equation}
\hat{X}_i|\psi)=|\hat{X}_i\psi),
\end{equation}
and the angular momentum operators with actions defined adjointly:
\begin{equation}
\label{angmom1}
\hat{L}_i|\psi) = | \frac{\hbar}{\theta}[\hat{X}_i,\psi])\equiv|\frac{\hbar}{\theta} ad_{\hat{X}_i}\psi).
\end{equation}
They satisfy the $SU(2)$ algebra commutation relations and obey the Leibnitz rule
\begin{equation} \label{li_leibnitz}
\hat{L}_i |\hat{\psi}\hat{\phi})= |\hat{\psi} (\hat{L}_i \hat{\phi})) + |(\hat{L}_i \hat{\psi})\hat{\phi}).
\end{equation}

In what follows it will turn out to be very useful to have a concrete realisation of the above Hilbert spaces and operators.  A particularly useful realisation is the Schwinger realisation of $SU(2)$.  Introducing the boson creation and annihilation operators $a^\dagger_\alpha,a_\alpha$, $\alpha=1,2$, this reads \cite{press}
\begin{eqnarray}
\hat{X}_i &&= \frac{\theta}{2} a_\alpha^\dagger \sigma_{\alpha\beta}^i a_\beta\quad ({\rm normal\;ordered}) \\
\hat{X}_i' &&= -\frac{\theta}{2} a_\alpha \sigma_{\alpha\beta}^i a_\beta^\dagger \quad ({\rm anti-normal\;ordered})
\end{eqnarray}
with $\sigma^i$ Pauli matrices.  Configuration space ${\cal H}_c$ then becomes the boson Fock space
\begin{equation}
\label{BFS}
{\cal H}_c={\rm span}\{|n_1,n_2\rangle=\frac{1}{\sqrt{n_1!n_2!}}(a_1^\dagger)^{n_1}(a_2^\dagger)^{n_2}|0,0\rangle\}.
\end{equation}
Since the Casimir is given by
\begin{equation}
\label{radcoord}
\hat{X}^2 = \hat{X}^{'2} = \hat{R} (\hat{R} + \theta), \quad \quad \hat{R} = \frac{\theta}{2} \hat{N}, \quad \quad \hat{N} = a_\alpha^\dagger a_\alpha,
\end{equation}
it is clear that each $SU(2)$ representation occurs precisely once, as desired.
The quantum Hilbert space ${\cal H}_q$ is then identified with the space of states
\begin{eqnarray}
\label{qhilbert}
{\cal H}_q=\left\{\hat{\psi}=\!\!\!\!\!\!\!\!\sum_{m_1,m_2,n_1,n_2 = 0}^{\infty} \!\!\!\!\!\!\!\!c_{m_1m_2n_1n_2} \left(a_1^\dagger\right)^{m_1} \left(a_2^\dagger\right)^{m_2} a_1^{n_1} a_2^{n_2}: m_1+m_2-n_1-n_2 = 0,\,{\rm tr_c}\left(\psi^\dagger\left(\hat{R}+\frac{\theta}{2}\right)^{1/2}\psi\right)<\infty\right\},\nonumber\\
\end{eqnarray}
equipped with the inner product (see (\ref{innprod_X}))
\begin{equation}
\label{innprod}
(\psi|\phi) = 8\pi\theta^2 \, {\rm tr_c}\left(\psi^\dagger\left(\hat{R} + \frac{\theta}{2}\right)\phi\right).
\end{equation}
Note that 
\begin{equation} \label{comm_R_psi}
[\hat{R}, \hat{\psi}] = 0
\end{equation}
as required for operators generated by the coordinates only. In what follows, we denote the basis elements of ${\cal H}_c$ by $|n_1,n_2\rangle=|j,m\rangle$, where $j=\frac{1}{2}(n_1+n_2)$, $m=\frac{1}{2}(n_1-n_2)$ and the elements of ${\cal H}_q$ by $|\psi)$.
Here $|n_1,n_2\rangle$ and $|j,m\rangle$ denotes the same state of ${\cal H}_c$ written in the notations of Fock and $SU(2)$ representations, respectively.

As described above, observables are identified with self-adjoint operators on ${\cal H}_q$, the most important ones being the angular momentum operators $\hat L_i$ with action defined adjointly as in (\ref{angmom1}) as well as: 
\begin{eqnarray}
\label{angmom}
\hat{L}^\prime_i|\psi) = |\frac{\hbar}{\theta} ad_{\hat{X}^\prime_i}\psi).
\end{eqnarray}
Note that $\hat{L}^2 = \hat{L}^{'2}$.  For later reference, it is useful to have an explicit form for the ladder operators:
\begin{equation} \label{l-}
\hat{L}_+\hat{\psi} = \left(\hat{L}_1+i\hat{L}_2\right)\hat{\psi} = \hbar\left(a_1^{\dagger}[a_2,\hat{\psi}]+[a_1^{\dagger},\hat{\psi}]a_2\right), \quad \quad 
\hat{L}_-\hat{\psi} = \left(\hat{L}_1-i\hat{L}_2\right)\hat{\psi}= \hbar\left(a_2^{\dagger}[a_1,\hat{\psi}]+[a_2^{\dagger},\hat{\psi}]a_1\right).
\end{equation}

It should be clear that there is a close connection between the Schwinger representation of the 3D fuzzy space and the Hopf fibration discussed earlier. Indeed, the connection follows quite straightforwardly through the introduction of coherent states on Boson Fock space introduced in (\ref{BFS}).  The most important features of this correspondence that we need are the following:  The operator $\frac{1}{2}\left(\bar{z}_\alpha\frac{\partial}{\partial \bar{z}_\alpha} + z_\alpha \frac{\partial}{\partial z_\alpha}\right)$ in $\mathds{C}^2$ corresponds to two different choices in the non-commutative case
\begin{eqnarray} \label{Delta_hat}
\hat{\Delta} \hat{\psi} &&= \frac{1}{2}\left(a_\alpha^\dagger[a_\alpha, \hat{\psi}] - [a_\alpha^\dagger, \hat{\psi}]a_\alpha\right)\quad{\rm (normal\;ordered)},\\
 \label{Delta_hat_prime}
\hat{\Delta}' \hat{\psi}&& = \frac{1}{2}\left([a_\alpha, \hat{\psi}]a_\alpha^\dagger - a_\alpha[a_\alpha^\dagger, \hat{\psi}] \right)\quad{\rm (anti-normal\;ordered)}.
\end{eqnarray}
These operators are related by
\begin{equation}
\hat{\Delta} - \hat{\Delta}' = ad_{a_\alpha^\dagger} ad_{a_\alpha} = ad_{a_\alpha} ad_{a_\alpha^\dagger} = [\hat{\Delta}, \hat{\Delta}'].
\end{equation}
When applied to a product of operators they give
\begin{equation} \label{delta_leibnitz}
\hat{\Delta}\left(\hat{\psi}\hat{\phi}\right) = \hat{\psi} \left(\hat{\Delta}\hat{\phi}\right) + \left(\hat{\Delta}\hat{\psi}\right) \hat{\phi} + \left[a_\alpha^\dagger, \hat{\psi}\right]\left[a_\alpha,\hat{\phi}\right], \quad \quad 
\hat{\Delta}' \left(\hat{\psi}\hat{\phi}\right) = \hat{\psi} \left(\hat{\Delta}' \hat{\phi}\right) + \left(\hat{\Delta}' \hat{\psi}\right) \hat{\phi} - \left[a_\alpha, \hat{\psi}\right]\left[a_\alpha^\dagger,\hat{\phi}\right].
\end{equation}
Furthermore, these operators commute with the angular momentum operators (\ref{angmom1}) and (\ref{angmom}) and can therefore at most involve (functions of) the radial coordinate $\hat{R}$ and the $\hat{L}^2$.  They are therefore the closest non-commutative analogues to the radial derivative introduced in the commutative case.  Indeed, when applied to an operator depending only on $\hat{R}$ they give
\begin{equation} \label{delta_psi_r}
\hat{\Delta}\hat{\psi}(\hat{R}) = \frac{2}{\theta} \hat{R} \left(\hat{\psi}(\hat{R}) - \hat{\psi}\left(\hat{R}-\frac{\theta}{2}\right)\right), \quad \quad 
\hat{\Delta}' \hat{\psi}(\hat{R}) = \frac{2}{\theta} \left(\hat{R}+\theta\right) \left(\hat{\psi}\left(\hat{R}+\frac{\theta}{2}\right) - \hat{\psi}(\hat{R})\right).
\end{equation}
Also note from (\ref{comm_R_psi}) that (\ref{Delta_hat}) and (\ref{Delta_hat_prime}) can be reduced to
\begin{equation}
\hat{\Delta} \hat{\psi} = a_\alpha^\dagger[a_\alpha, \hat{\psi}], \quad \quad
\hat{\Delta}' \hat{\psi} = [a_\alpha, \hat{\psi}]a_\alpha^\dagger. 
\end{equation}

We can now proceed to write the Hamiltonian for a particle moving in 3D fuzzy space.  We start with the free particle Hamiltonian, which is given by
\begin{equation} \label{Hamiltonian}
\hat{H}_0 = \frac{\hbar^2}{\theta m_0 \left(\hat{R}+\frac{\theta}{2}\right)} (\hat{\Delta} - \hat{\Delta}') 
=  \frac{\hbar^2}{\theta m_0 \left(\hat{R}+\frac{\theta}{2}\right)} ad_{a_\alpha^\dagger} ad_{a_\alpha}.
\end{equation}
This Hamiltonian is hermitian with respect to the inner product (\ref{innprod}) as can be easily verified.  It can also be easily checked that it is a non-negative operator. It is important to note that the operators $\hat{\Delta}$ and $\hat{\Delta}'$ are not separately hermitian with respect to this inner product, but only the combination introduced above.  Finally, it can be checked that the angular momentum operators $\hat{L}_i$ are also hermitian with respect to this inner product.
We now proceed to solve the time independent free particle Schr\"odinger equation: 
\begin{equation} \label{Schrodinger_eqn}
\hat{H}_0 \hat{\psi} = E \hat{\psi}
\end{equation}

\section{Solving the fuzzy free particle Schr\"odinger equation}
\label{three}
We start by introducing the non-commutative analogues of the mixed spherical harmonics introduced for the commutative case in section \ref{one}:
\begin{eqnarray}
\hat{\mathcal{Y}}_{lm} = c_{lm} \hat{L}_-^{l-m} \hat{\mathcal{Y}}_{ll}, \quad \quad \hat{\mathcal{Y}}_{ll} = \left(a_1^\dagger\right)^l a_2^l \label{ylm},\\
\hat{\mathcal{Y}}'_{lm} = c_{lm} \hat{L}_-^{\prime l-m} \hat{\mathcal{Y}}'_{ll}, \quad \quad \hat{\mathcal{Y}}'_{ll} = \hat{\mathcal{Y}}_{l,-l} = (-1)^l \left(a_2^\dagger\right)^l a_1^l.\label{ylm'}
\end{eqnarray}
They satisfy
\begin{eqnarray}
\hat{L}^2 \hat{\mathcal{Y}}_{lm} = l(l+1) \hbar^2 \hat{\mathcal{Y}}_{lm}, \quad \quad \hat{L}_3 \hat{\mathcal{Y}}_{lm} = m \hbar \hat{\mathcal{Y}}_{lm}, \\
\hat{L}^{\prime 2} \hat{\mathcal{Y}}'_{lm} = l(l+1) \hbar^2 \hat{\mathcal{Y}}'_{lm}, \quad \quad \hat{L}'_3 \hat{\mathcal{Y}}'_{lm} = m \hbar \hat{\mathcal{Y}}'_{lm}.
\end{eqnarray}
The action of $\hat{\Delta}$ and $\hat{\Delta}'$ on them is given by
\begin{eqnarray} 
\hat{\Delta} \hat{\mathcal{Y}}_{lm} = \hat{\Delta}' \hat{\mathcal{Y}}_{lm} = l \hat{\mathcal{Y}}_{lm}, \quad \quad
\hat{\Delta} \hat{\mathcal{Y}}'_{lm} = \hat{\Delta}' \hat{\mathcal{Y}}'_{lm} = l \hat{\mathcal{Y}}'_{lm},
&\quad \Rightarrow \quad&
(\hat{\Delta} - \hat{\Delta}') \hat{\mathcal{Y}}_{lm}
= (\hat{\Delta} - \hat{\Delta}') \hat{\mathcal{Y}}'_{lm}
= 0. \label{Delta_Ylm}
\end{eqnarray}
It follows that
\begin{equation}
\hat{H}_0 \hat{\mathcal{Y}}_{lm} = \hat{H}_0 \hat{\mathcal{Y}}'_{lm}  = 0,
\end{equation}
in line with (\ref{H0_mixedYlm}).
Another important feature of the non-commutative mixed spherical harmonics is the following:
\begin{equation} \label{ylm_kernel_1}
\hat{\mathcal{Y}}_{lm} |n_1,n_2\rangle = 0 \quad \quad {\rm for} \,\, n=n_1+n_2<l.
\end{equation}
This result, as well as a more extensive discussion of the non-commutative mixed spherical harmonics can be found in appendix \ref{A}. \\

The next step is to write the wave function, as is usually done, as the product of a radial function and $\hat{\mathcal{Y}}_{lm}$ 
\begin{equation} \label{psi_hR_ylm}
\hat{\psi}_{k,l,m} = \hat{h}_{k,l}(\hat{R}) \, \hat{\mathcal{Y}}_{lm}
\end{equation}
with $k$ given in (\ref{eqn_gr_comm}).
Substituting it in the Schr\"{o}dinger equation (\ref{Schrodinger_eqn}) gives the equation for $\hat{h}_{k,l}(\hat{R})$:
\begin{equation} \label{eqn_hr}
\left(\hat{R}-\frac{l\theta}{2}\right)\hat{h}_{k,l} \left(\hat{R}-\frac{\theta}{2}\right) 
- 2 \left(1-\frac{\theta^2 k^2}{8}\right) \left(\hat{R}+\frac{\theta}{2}\right)
\hat{h}_{k,l} (\hat{R}) 
+ \left(\hat{R}+\frac{(l+2)\theta}{2}\right)\hat{h}_{k,l} \left(\hat{R}+\frac{\theta}{2}\right) 
= 0.
\end{equation}
Here we have used (\ref{Hamiltonian}), (\ref{delta_leibnitz}), (\ref{delta_psi_r}), (\ref{Delta_Ylm}), (\ref{adagger_f_a_y}) and (\ref{a_f_adagger_y}).
As the operator on the right hand side of (\ref{psi_hR_ylm}) vanishes when applied to states $|n_1,n_2\rangle $ with $n_1+n_2 < l$ (see (\ref{ylm_kernel})), the above constraint is required only in the domain of the configuration space spanned by the states $|n_1,n_2\rangle $ with $n_1+n_2 \geq l$.
Writing
 \begin{equation} \label{R_in_N}
  \hat{R}=\frac{\theta}{2}\hat{N}, \quad \quad \hat{h}_{k,l}(\hat{R}) = \hat{h}_{k,l}\left(\frac{\theta}{2}\hat{N}\right) \rightarrow \hat{h}_{k,l}(\hat{N}),
 \end{equation}
we get
\begin{equation} \label{eqn_hN}
\left(\hat{N}-l\right)\hat{h}_{k,l}(\hat{N}-1) 
- 2 \xi \left(\hat{N}+1\right)
\hat{h}_{k,l}(\hat{N}) 
+ \left(\hat{N}+l+2\right)\hat{h}_{k,l}(\hat{N}+1) 
= 0
\end{equation}
with
\begin{equation} \label{xi}
\xi = 1-\frac{\theta^2 k^2}{8} = 1-\frac{\theta^2 m_0 E}{4\hbar^2}.
\end{equation}
The projection operator
\begin{equation}
\hat{P}_j = \sum_{m=-j}^{j}|j,m\rangle\langle j,m| = \sum_{n_1=0}^{n}|n_1,n-n_1\rangle\langle n_1,n-n_1| 
\end{equation}
projects, in the configuration Hilbert space ${\cal H}_c$, onto the eigenspace of the number operator $\hat{N}$ with eigenvalue $n$.
Recalling the relation (\ref{radcoord}) between the radial coordinate $\hat{R}$ and $\hat{N}$, this operator has the physical meaning of projecting onto a shell of fixed radius in configuration space. 
A function of $\hat{N}$ now can be expanded as
\begin{eqnarray}
 \hat{h}(\hat{N}) &=& \sum_{n_1,n_2=0}^{\infty}h(n_1+n_2)|n_1,n_2\rangle \langle n_1.n_2| \nonumber \\
&=& \sum_{j=0,\frac{1}{2},1,...}h(2j)\hat{P}_j \nonumber \\
&=&  \sum_{n=0}^{\infty} h(n)\hat{P}_{n/2} \label{hN_hn}.
\end{eqnarray}
Substituting this form for $\hat{h}_{k,l}(\hat{N})$ in (\ref{eqn_hN}) and using the properties of the projection operators we get, for $n\geq l$, the difference equation:
\begin{equation} \label{eqn_hn}
(n-l) \, h_{k,l}(n-1) - 2 \xi (n+1) \, h_{k,l}(n) + (n+l+2) \, h_{k,l}(n+1) = 0.
\end{equation}
\\

Let us first solve (\ref{eqn_hn}) for the case of $l=0$, in which case it reads: 
\begin{equation} \label{eqn_hn_l=0}
n \, h_{k,0}(n-1) - 2 \xi (n+1) \, h_{k,0}(n) + (n+2) \, h_{k,0}(n+1) = 0, \quad \quad n\geq 0.
\end{equation}
The solutions are given by
\begin{equation} \label{sol_hn_l=0}
h_{k,0}(n) = \frac{\mu^n}{n+1}
\end{equation}
where $\mu$ satisfies
\begin{equation} \label{eqn_mu}
\mu^2-2\xi\mu+1 = 0,
\end{equation}
and is explicitly given by
\begin{equation} \label{mu_in_xi}
\mu = \xi \pm \sqrt{\xi^2 - 1}.
\end{equation}
Note that the solution (\ref{sol_hn_l=0}) satisfies the equation (\ref{eqn_hn_l=0}) for $n\geq 1$, but not for $n=0$.
A general solution  (for $\xi \neq \pm 1$) is given by
\begin{equation} \label{sol_hn_general_l=0}
h_{k,0}(n) = A_1 h_{k,0}^{(1)}(n) + A_2 h_{k,0}^{(2)}(n),
\end{equation}
with
\begin{equation} \label{sol_h_alpha_n_l=0}
h_{k,0}^{(\alpha)}(n) = \alpha_0 \frac{\mu_{\alpha}^n}{n+1}, \quad \quad \alpha=1,2,
\end{equation}
$\mu_1,\mu_2$ being the two solutions of (\ref{eqn_mu}) given by (\ref{mu_in_xi}).
$\alpha_0$, an arbitrary constant, has been introduced so that the solution matches exactly the solution for general integral $l$, which is discussed next, when we set $l=0$. 
The solution (\ref{sol_hn_general_l=0}) satisfies (\ref{eqn_hn_l=0}) even for $n=0$, i.e.,
\begin{equation}
h_{k,0}(1) = \xi h_{k,0}(0),
\end{equation}
provided the constants $A_1$ and $A_2$ are related as
\begin{equation} \label{relation_A1_A2_1}
A_2 = -\frac{h_{k,0}^{(1)}(1) - \xi h_{k,0}^{(1)}(0)}{h_{k,0}^{(2)}(1) - \xi h_{k,0}^{(2)}(0)} A_1
= -\frac{\mu_1 - 2\xi}{\mu_2 - 2\xi} A_1
= -\frac{\mu_2}{\mu_1} A_1.
\end{equation}
\\

Next we solve (\ref{eqn_hn}) for any positive integer $l$. To do so we use a suitable adaptation of the standard technique described in \cite{Jagerman}.
Replacing $n\rightarrow n-1$ in the (\ref{eqn_hn}) we get, for $n\geq l+1$,
\begin{equation} \label{eqn_hn_1}
(n-l-1) \, h_{k,l}(n-2) - 2 \xi n \, h_{k,l}(n-1) + (n+l+1) \, h_{k,l}(n) = 0.
\end{equation}
Let us introduce the notations
\begin{eqnarray}
E u(n) &=& u(n+1), \nonumber\\
E^{-1}u(n) &=& u(n-1),\nonumber\\
 \delta u(n)= &\left(E-1\right)u(n)& = u(n+1)-u(n),\nonumber \\
\delta_{(-1)}u(n)= &\left(1-E^{-1}\right)u(n)& = u(n)-u(n-1),
\end{eqnarray}
to define operators $\rho$ and $\pi$ as
\begin{eqnarray}
\rho u(n)=(n-l)E^{-1}u(n)=(n-l)u(n-1),\nonumber\\ 
\pi u(n)=(n-l)\delta_{(-1)}u(n)=(n-l)\left[u(n)-u(n-1)\right].
\end{eqnarray}

For any positive integer $k$ we then have
\begin{equation}
\rho^k u(n)=\frac{\Gamma(n-l+1)}{\Gamma(n-l-k+1)}E^{-k}u(n)=\frac{\Gamma(n-l+1)}{\Gamma(n-l-k+1)}u(n-k), 
\end{equation}
and we define $\rho^k$ for any complex $k$ through analytic continuation. Setting
\begin{equation} \label{gn_vn}
 h_{k,l}(n)=\mu^n \, v_{k,l}(n)
\end{equation}
in (\ref{eqn_hn_1}) yields
\begin{equation} \label{eqn_vn}
(n-l-1) \, v_{k,l}(n-2) - 2 \xi \mu n \, v_{k,l}(n-1) + \mu^2 (n+l+1) \, v_{k,l}(n) = 0.
\end{equation}
Multiplying with $(n-l)$ and writing $(n-l) \, v_{k,l}(n-1)$ and $(n-l)(n-l-1) \, v_{k,l}(n-2)$ as $\rho v_{k,l}(n)$ and $\rho^2 v_{k,l}(n)$, respectively, the above equation can be recast  in the form
\begin{equation}
 \left[\rho^2-2\mu\xi n\rho+\mu^2(n-l)(n+l+1)\right] v_{k,l}(n)=0.
\end{equation}
Observing that
\begin{equation}
\left(\pi+\rho+l\right)u(n)=nu(n),
\end{equation}
one can replace $n$ by $\pi+\rho+l$ and using
\begin{equation}
\rho\pi=\pi\rho-\rho,
\end{equation}
leads to
\begin{equation} \label{eqn_vn_rho_pi}
\left[(\mu^2-2\xi\mu+1)\rho^2 + 2\mu(\mu-\xi)\pi\rho + 2l\mu(\mu-\xi)\rho +
\mu^2\pi \left(\pi+2l+1\right)\right] v_{k,l}(n) = 0.
\end{equation}
As $\mu$ satisfies (\ref{eqn_mu}), this equation becomes
\begin{equation} \label{eqn_vn_fpi}
 \left[f_1(\pi)\rho+f_0(\pi)\right] v_{k,l}(n)=0,
\end{equation}
where
\begin{equation}
f_1(\pi) = 2\mu(\mu-\xi)(\pi+l), \quad \quad f_0(\pi) = \mu^2 \pi (\pi + 2l + 1).
\end{equation}
Consider the factorial functions
\begin{equation} \label{factorial_function}
\tilde{\rho}^{k}=\rho^{k}1 = \frac{\Gamma(n-l+1)}{\Gamma(n-l-k+1)}.
\end{equation}
For different integral values of $k$ we have
\begin{equation} 
\tilde{\rho}^{k} = \left\{
\begin{array}{cl}
(n-l)(n-l-1)...(n-l-k+1) & {\rm for} \,\, k>0, \nonumber\\
1 & {\rm for} \,\, k=0, \nonumber\\
\frac{1}{(n-l+1)(n-l+2)...(n-l-k)} & {\rm for} \,\, k<0. 
\end{array}
\right.
\end{equation}
Again these factorial functions are defined for complex $k$ by analytic continuation. 
The following relations can be readily checked:
\begin{eqnarray}
\rho^m \tilde{\rho}^k& = &\tilde{\rho}^{m+k},\nonumber\\
\tilde{\rho}^0 &=& 1,\nonumber\\
 \pi \tilde{\rho}^k&=&k\tilde{\rho}^k,\nonumber\\
 \pi^m \tilde{\rho}^k&=&k^m\tilde{\rho}^k,\nonumber\\
 f(\pi) \tilde{\rho}^k&=&f(k)\tilde{\rho}^k,
\end{eqnarray}
where $f(\pi)$ is any polynomial in the operator $\pi$.
Taking the solution to be an inverse factorial series
\begin{equation}
v_{k,l}(n)=\sum_{s=0}^{\infty}\alpha_s\tilde{\rho}^{k-s},
\end{equation}
substituting in (\ref{eqn_vn_fpi}) and using the above results, yield
\begin{equation}
 \alpha_0f_1(k+1)\tilde{\rho}^{k+1}+\sum_{s=0}^{\infty}\left\{\alpha_{s+1}f_1(k-s)+\alpha_{s}f_0(k-s)\right\}\tilde{\rho}^{k-s}=0.
\end{equation}
The $\tilde{\rho}^k$'s for different $k$'s are linearly independent. Indeed, the Casorati's determinant for two of them are given by
\begin{eqnarray}
W(\tilde{\rho}^k,\tilde{\rho}^{k^\prime}) &=& \left|
\begin{array}{cc}
 \tilde{\rho}^k & \tilde{\rho}^{k^\prime} \\
\tilde{\rho}^{k+1} & \tilde{\rho}^{{k^\prime}+1}
\end{array}
\right| \nonumber \\
&=& (k-k^\prime)\tilde{\rho}^{k}\tilde{\rho}^{{k^\prime}} \nonumber \\
&\neq& 0 \mbox{\,\,\,\,\rm{for }}k\neq k^{\prime}.
\end{eqnarray}
Equating the coefficients of different $\tilde{\rho}^k$'s yields
\begin{eqnarray}
 \alpha_0f_1(k+1) &=& 0 \mbox{\,\,\,\,\,\,\rm{; \, indicial equation}}, \\
\alpha_{s+1}f_1(k-s)+\alpha_{s}f_0(k-s) &=& 0.
\end{eqnarray}
The indicial equation gives
\begin{equation}
k=-l-1 \quad \quad \mbox{{\rm for \,\,}}  \xi \neq \pm 1,
\end{equation}
while the other equation gives
\begin{equation}
 \alpha_{s+1}=p(s)\alpha_s, \quad \quad \quad
 p(s) = -\frac{f_0(k-s)}{f_1(k-s)}
= \frac{\mu (l+s+1)(s-l)}{2(\mu-\xi)(s+1)}.
\end{equation}
$\alpha_s$ (for $s>0$) is given by
\begin{equation}
 \alpha_s = \alpha_0\prod_{r=0}^{s-1}p(r)  = \left(\frac{\mu}{2(\xi-\mu)}\right)^s \frac{(l-s+1)(l-s+2)...(l+s)}{s!} \alpha_0.
\end{equation}
Note that $\alpha_s$ vanishes for all $s>l$. Thus
\begin{equation}
\alpha_s = \left\{
\begin{array}{cc}
 \left(\frac{\mu}{2(\xi-\mu)}\right)^s \frac{(l+s)!}{(l-s)!s!} \alpha_0 & {\rm for} \,\, 0\leq s \leq l,\nonumber\\
 0 & {\rm for} \,\, s>l.
\end{array}
\right.
\end{equation}
We therefore get
\begin{equation} \label{sol_hn}
h_{k,l}(n) = \mu^n \sum_{s=0}^{l} \alpha_0 \left(\frac{\mu}{2(\xi-\mu)}\right)^s \frac{(n-l)! \,\, (l+s)!}{(n+s+1)!\,\,(l-s)!\,\,s!}.
\end{equation}
Note that this solution for $h_{k,l}(n)$ is well defined for $n\geq l$.
On the other hand it satisfies (\ref{eqn_hn_1}) only for $n>l+1$, but not for $n=l+1$.
A general solution is given by
\begin{equation} \label{sol_hn_general}
h_{k,l}(n) = A_1 h_{k,l}^{(1)}(n) + A_2 h_{k,l}^{(2)}(n),
\end{equation}
where
\begin{equation} \label{sol_h_alpha_n}
h_{k,l}^{(\alpha)}(n) = \mu_\alpha^n \sum_{s=0}^{l} \alpha_0 \left(\frac{\mu_\alpha}{2(\xi-\mu_\alpha)}\right)^s \frac{(n-l)! \,\, (l+s)!}{(n+s+1)! \,\, (l-s)! \,\, s!}, \quad \quad \alpha=1,2,
\end{equation}
and $\mu_1,\mu_2$ are the two solutions of (\ref{eqn_mu}) given by (\ref{mu_in_xi}).
The solution (\ref{sol_hn_general}) satisfies (\ref{eqn_hn_1}) even for $n=l+1$, i.e.
\begin{equation} \label{bc_origin}
h_{k,l}(l+1) = \xi h_{k,l}(l),
\end{equation}
if the constants $A_1$ and $A_2$ are related by
\begin{equation} \label{relation_A1_A2}
A_2 = -\frac{\xi h_{k,l}^{(1)}(l) - h_{k,l}^{(1)}(l+1)}{\xi h_{k,l}^{(2)}(l) - h_{k,l}^{(2)}(l+1)} A_1.
\end{equation}

As a consistency check, we investigate the commutative limit of these solutions.  This can be done on two levels, firstly one can check that the radial equation (\ref{eqn_hr}) has the appropriate commutative limit and, secondly, one can check that the solutions above reduce to the commutative solutions.  let us first consider the commutative limit of the radial equation (\ref{eqn_hr}).  Expanding $\hat{h}_{k,l} \left(\hat{R}-\frac{\theta}{2}\right)$ and $\hat{h}_{k,l} \left(\hat{R}+\frac{\theta}{2}\right)$ in $\theta$ yields
\begin{eqnarray}
\hat{h}_{k,l} \left(\hat{R}-\frac{\theta}{2}\right) &=& h_{k,l}(r) - \frac{\theta}{2} h'_{k,l}(r) + \frac{\theta^2}{8} h''_{k,l}(r) + ...,\nonumber \\
 \hat{h}_{k,l} \left(\hat{R}+\frac{\theta}{2}\right) &=& h_{k,l}(r) + \frac{\theta}{2} h'_{k,l}(r) + \frac{\theta^2}{8} h''_{k,l}(r) + ...\;.
\end{eqnarray}
Substituting in (\ref{eqn_hr}) indeed yields  the commutative radial equation (\ref{eqn_hr_comm}).
Next we turn to the commutative limit of the solutions.  Substituting (\ref{xi}) in (\ref{mu_in_xi}) yields
\begin{equation}
\mu = 1- \frac{\theta^2 k^2}{8} \pm \frac{\theta k}{2} \sqrt{\frac{\theta^2 k^2}{16}-1} 
\, = \, 1- \frac{\theta^2 m_0 E}{4\hbar^2} \pm \frac{\theta}{\hbar} \left(\frac{m_0 E}{2}\right)^{1/2} \sqrt{\frac{\theta^2 m_0 E}{8\hbar^2}-1}
\end{equation}

Upon setting (see (\ref{R_in_N})) $r = n \frac{\theta}{2}$ one readily finds:
\begin{eqnarray}
\lim_{\theta\rightarrow 0} \mu^n &=& \exp\{\displaystyle{\lim_{\theta\rightarrow 0}}\, n\ln \mu\} = e^{\pm ikr}, \\
\frac{\mu}{\xi-\mu}& \approx & \pm \frac{2i}{\theta k} + {\rm higher \,\, orders \,\, in} \,\, \theta, \\
\frac{(n-l)!}{(n+s+1)!} &=& \frac{1}{(n+s+1)(n+s)...(n-l+1)} \nonumber \\
&=& \frac{1}{\left(\frac{2r}{\theta}+s+1\right)\left(\frac{2r}{\theta}+s\right)...\left(\frac{2r}{\theta}-l+1\right)} \nonumber \\
& \approx & \left(\frac{\theta}{2r}\right)^{l+s+1} + {\rm higher \,\, orders \,\, in} \,\, \theta,
\end{eqnarray}
from which it follows that (\ref{sol_hn}) reduces in the commutative limit to
\begin{equation}
h_{k,l}(r) = \left(\frac{\theta}{2r}\right)^{l+1} \alpha_0^{\theta\rightarrow 0} e^{\pm ikr} y_l\left(\pm \frac{i}{kr}\right),
\end{equation}
where $ \alpha_0^{\theta\rightarrow 0}$ denotes the commutative limit of $\alpha_0$ and the Bessel polynomials $y_l(x)$ appearing here are given in (\ref{bessel_expansion}).
Upon choosing $\alpha_0$ such that its commutative limit satisfies
\begin{equation} \label{alpha_0_limit}
\theta^{l+1} \alpha_0^{\theta\rightarrow 0} = \frac{2^{l+1}}{k},
\end{equation}
the solution goes to the correct limit as given in (\ref{hr_bessel_polynomial}).

\section{Interpreting the spectrum of the fuzzy free particle}
\label{four}
In this section we analyse the spectrum of the fuzzy free particle in more detail.  For this purpose, it is convenient to define a dimensionless energy
\begin{equation} \label{epsilon_E}
\varepsilon = \frac{\theta^2 m_0 E}{8\hbar^2} = \frac{\theta^2 k^2}{16},
\end{equation}
in terms of which (\ref{xi}) and (\ref{mu_in_xi}) read
\begin{equation}
\xi = 1-2\varepsilon,\quad\mu = 1-2\varepsilon \pm 2\sqrt{\varepsilon (\varepsilon -1)}.
\end{equation}
As already pointed out in section \ref{two}, the free particle Hamiltonian is non-negative and subsequently the energy $E$ and hence $\varepsilon$ are also non-negative as indicated in (\ref{epsilon_E}). \\

We start by considering the range of energies $0<\varepsilon<1$.   In this case $\mu$ is complex:
\begin{equation}
\mu = 1-2\varepsilon \pm 2i\sqrt{\varepsilon (1-\varepsilon)}.
\end{equation}
Since $-1<1-2\varepsilon<1$, one can always choose an angle $0<\varphi<\pi$ such that
\begin{equation} \label{sinphi_cosphi}
\xi = 1-2\varepsilon = \cos\varphi, \quad \quad \sqrt{1-\xi^2} = 2\sqrt{\varepsilon (1-\varepsilon)} = \sin\varphi.
\end{equation} 
In terms of this, we get
\begin{equation}
\mu_1 = e^{i\varphi}, \quad \quad \mu_2 = e^{-i\varphi}, \quad \quad \quad
\varepsilon = \sin^2 \left(\frac{\varphi}{2}\right)
\end{equation}
and
\begin{equation} \label{k_phi}
k = \frac{1}{\hbar} \sqrt{2m_0 E} = \frac{4}{\theta} \sqrt{\varepsilon} = \frac{4}{\theta} \sin\left(\frac{\varphi}{2}\right).
\end{equation}
The solution (\ref{sol_h_alpha_n}) becomes
\begin{eqnarray}
h_{\varphi,l}^{(1)}(n) &=&\frac{1}{2} \left(\frac{2}{\theta}\right)^{l} \csc \left(\frac{\varphi}{2}\right) e^{in\varphi} \sum_{s=0}^{l} \frac{e^{is\left(\varphi+\frac{\pi}{2}\right)}}{(2 \sin\varphi)^s} \frac{(n-l)! \, (l+s)!}{(n+s+1)! \, (l-s)! \, s!} \label{sol_hl1_E<1},\nonumber \\
h_{\varphi,l}^{(2)}(n) &=& h_{\varphi,l}^{(1)*}(n) = \frac{1}{2} \left(\frac{2}{\theta}\right)^{l} \csc \left(\frac{\varphi}{2}\right) e^{-in\varphi}  \sum_{s=0}^{l} \frac{e^{-is\left(\varphi+\frac{\pi}{2}\right)}}{(2 \sin\varphi)^s} \frac{(n-l)! \, (l+s)!}{(n+s+1)! \, (l-s)! \, s!} \label{sol_hl2_E<1}.
\end{eqnarray}
Here we have chosen $\alpha_0 = \left(\frac{2}{\theta}\right)^{l+1}\frac{1}{k} = \frac{1}{2} \left(\frac{2}{\theta}\right)^{l} \csc \left(\frac{\varphi}{2}\right)$ (see (\ref{alpha_0_limit}) and (\ref{k_phi})).
Also the quantum number $k$ has been replaced by $\varphi$ as they are related by the bijection (\ref{k_phi}) for the energy range $0<\varepsilon<1$ with $0<\varphi<\pi$.
The property $h_{\varphi,l}^{(2)}(n) = h_{\varphi,l}^{(1)*}(n)$ simplifies the relation (\ref{relation_A1_A2}) to
\begin{equation} \label{A2_A1_E<1}
A_2 =  -\frac{\cos\varphi \, h_{\varphi,l}^{(1)}(l) - h_{\varphi,l}^{(1)}(l+1)}{\cos\varphi \, h_{\varphi,l}^{(1)*}(l) - h_{\varphi,l}^{(1)*}(l+1)} A_1
= -e^{2i\delta_{\varphi,l}} A_1, \quad \quad \quad
\delta_{\varphi,l} = {\rm arg}\left[\cos\varphi \,\, h_{\varphi,l}^{(1)}(l) - h_{\varphi,l}^{(1)}(l+1)\right].
\end{equation}
The full fuzzy free particle solution can now be written down: 
\begin{equation} \label{psi_hn_ylm}
\hat{\psi}_{\varphi,l,m} = \hat{h}_{\varphi,l}(\hat{N}) \, \hat{\mathcal{Y}}_{lm},
\end{equation}
where $\hat{h}_{\varphi,l}(\hat{N})$ is expanded as in (\ref{hN_hn}) with the coefficients being given by
\begin{equation} \label{sol_hn_general_withO}
h_{\varphi,l}(n) = A_1 \left[h_{\varphi,l}^{(1)}(n) - e^{2i\delta_{\varphi,l}} \, h_{\varphi,l}^{(1)*}(n)\right], \quad \quad n\geq l.
\end{equation}
Note that in order to get the action of $\hat{\psi}_{\varphi,l,m}$ on an arbitrary Fock state $|n_1,n_2\rangle$, one does not need to know $h_{\varphi,l}(n)$ for $n<l$, as $\hat{\mathcal{Y}}_{lm}$ (and hence $\hat{\psi}_{\varphi,l,m}$) annihilates the states $|n_1,n_2\rangle$ with $n=n_1+n_2<l$.
Taking $e^{i\delta_{\varphi,l}}$ out of the bracket and absorbing it into the arbitrary coefficient we get for $n\geq l$
\begin{equation} \label{sol_hn_E<1_nonsingular}
h_{\varphi,l}(n) = A\,\, \Im \left[e^{-i\delta_{\varphi,l}} \, h_{\varphi,l}^{(1)}(n)\right] 
= \frac{A}{2} \left(\frac{2}{\theta}\right)^{l} \csc \left(\frac{\varphi}{2}\right) \sum_{s=0}^{l} \frac{\sin \left[n\varphi + s\left(\varphi+ \frac{\pi}{2}\right) - \delta_{\varphi,l}\right]}{(2 \sin\varphi)^s} \frac{(n-l)! \, (l+s)!}{(n+s+1)!\, (l-s)!\, s!},
\end{equation}
where
\begin{equation}
A = 2i e^{i\delta_{\varphi,l}} A_1,
\end{equation}
and $\Im (z)$ denotes the imaginary part of the complex number $z$. As the $h_{\varphi,l}(n)$ in (\ref{sol_hn_E<1_nonsingular}) is pure real (up to the complexity of the normalization constant $A$), it represents a standing wave.
The constant $A$ is determined by requiring the wave function $\hat{\psi}_{\varphi,l,m}$ to be normalized in the following way
\begin{equation} \label{orthogonality_psi}
\left(\hat{\psi}_{\varphi, l, m} \, , \, \hat{\psi}_{\varphi', l', m'}\right)
= \delta (\varphi - \varphi') \, \delta_{ll'} \, \delta_{mm'}
\end{equation}
Note that the orthogonality of the wave functions is ensured by the hermiticity of the  Hamiltonian (\ref{Hamiltonian}). \\

For the purposes of a scattering and phase analysis, it is useful to find the asymptotic behaviour of these solutions. The leading asymptotic behaviour of the inverse factorial function for $n\rightarrow \infty$ is:
\begin{equation}
\frac{(n-l)!}{(n+s+1)!} = \frac{1}{(n-l+1)(n-l+2)...(n+s+1)} \rightarrow \frac{1}{n^{l+s+1}}, \quad \quad s \geq 0.
\end{equation}
Hence
\begin{equation}
\left. h_{\varphi,l}^{(1)}(n)\right|_{n\rightarrow \infty} = \frac{1}{2} \left(\frac{2}{\theta}\right)^{l} \csc \left(\frac{\varphi}{2}\right) \frac{e^{in\varphi}}{n^{l+1}},
\end{equation}
representing an outgoing spherical wave, while $h_{\varphi,l}^{(1)*}(n)$ represents an incoming spherical wave.
$h_{\varphi,l}(n)$ represents an incident incoming spherical wave of energy $E$ getting fully reflected at the origin to produce an outgoing spherical wave of the same energy with a phase difference of $2\delta_{\varphi,l}$. 
The energy dependence of the frequency of the spherical wave $\varphi$ is different from the commutative case, reflecting a modified dispersion relation. 
Modified dispersion relations have important  thermodynamic consequences that have been studied extensively (see for example \cite{Gregg:2008jb, Camacho:2007qy, Chandra:2011nj}).
Note that in the commutative limit (see (\ref{k_phi}))
\begin{equation}
\varphi |_{\theta\rightarrow 0} \sim \frac{\theta k}{2} \quad \quad \Rightarrow \quad \quad e^{in\varphi} \rightarrow e^{ikr},
\end{equation}
and subsequently that $h_{\varphi,l}^{(1)}(n)$ reduces to the appropriate outgoing spherical wave in the commutative limit:
\begin{equation}
\left. h_{\varphi,l}^{(1)}(n)\right|_{n\rightarrow \infty, \theta \rightarrow 0} = \frac{e^{ikr}}{k r^{l+1}}.
\end{equation}

Finally, for consistency, we also check the commutative limit of the phase difference. In this regard,  note that the phase difference between incoming and outgoing spherical waves for a free particle as given in (\ref{A2_A1_E<1}) depends on both the $\varphi-$value (and hence energy) and the $l-$value (i.e., the angular momentum) in contrast with the commutative case where it is given by $\frac{l\pi}{2}$ and does not depend on the energy or momentum.
Using (\ref{epsilon_E}) and (\ref{sinphi_cosphi}) we write
\begin{equation}
\cos\varphi = 1- \frac{\theta^2 k^2}{8}
\end{equation}
and expand $h_{\varphi,l}^{(1)}(l)$ and $h_{\varphi,l}^{(1)}(l+1)$ in $\theta$:
\begin{eqnarray}
h_{\varphi,l}^{(1)}(l+1) = h_{\varphi,l}^{(1)} \left(r= \frac{(l+1)\theta}{2}\right) &=& h_{\varphi,l}^{(1)}(r=0) \, + \, \frac{(l+1)\theta}{2} h_{\varphi,l}^{(1)\prime}(r=0) \, + \, \mathcal{O}(\theta^2),\nonumber \\
h_{\varphi,l}^{(1)}(l) = h_{\varphi,l}^{(1)} \left(r= \frac{l\theta}{2}\right) &=& h_{\varphi,l}^{(1)}(r=0) \, + \, \frac{l\theta}{2} h_{\varphi,l}^{(1)\prime}(r=0) \, + \,   \mathcal{O}(\theta^2),
\end{eqnarray}
to obtain
\begin{equation}
\cos\varphi \,\, h_{\varphi,l}^{(1)}(l) - h_{\varphi,l}^{(1)}(l+1)
= \,\, - \frac{\theta}{2} \, h_{\varphi,l}^{(1)\prime} (r=0) \, + \, \mathcal{O}(\theta^2).
\end{equation}
Using (\ref{bessel_expansion}) and (\ref{hr_bessel_polynomial}) one finds
\begin{equation}
\left. h_{\varphi,l}^{(1)\prime}(r)\right|_{r\rightarrow 0} \sim - \frac{(2l+1)!}{2^l k^{l+1} r^{2l+2} \, l!} e^{i\frac{l\pi}{2}}.
\end{equation}
The relation between the coefficients, i.e., (\ref{A2_A1_E<1}) in the commutative limit then becomes
\begin{eqnarray}
A_2 = - e^{il\pi} A_1, &\quad \Rightarrow \quad & 
\left. \delta_{\varphi,l}\right|_{\theta\rightarrow 0} = \frac{l\pi}{2}.
\end{eqnarray}
\\

Next we consider $\varepsilon = 0$ in which case $\xi = 1$ and the difference equation (\ref{eqn_hn_1}) becomes ($n\geq l+1$)
\begin{equation} \label{eqn_hn_E=0}
(n-l-1) \, h_{k,l}(n-2) - 2n \, h_{k,l}(n-1) + (n+l+1) \, h_{k,l}(n) = 0.
\end{equation}
Multiplying with $(n-l)$, writing $(n-l) \, h_{k,l}(n-1)$ and $(n-l)(n-l-1) \, h_{k,l}(n-2)$ as $\rho h_{k,l}(n)$ and $\rho^2h_{k,l}(n)$, respectively, replacing $n$ by $\pi+\rho+l$ and using $\rho\pi=\pi\rho-\rho$ yield
\begin{equation}
 \pi (\pi+2l+1) \, h_{k,l}(n) = 0.
\end{equation}
A general solution to the above equation is given by
\begin{equation} \label{hn_sol_E=0}
h_{k,l}(n) = A + B \tilde{\rho}^{-2l-1}  =
A + \frac{B}{(n-l+1)(n-l+2)...(n+l+1)} .
\end{equation}
The equation (\ref{eqn_hn_E=0}) for $n=l+1$ gives $h_{k,l}(l+1) = h_{k,l}(l)$, which requires $B=0$.  Subsequently we get $h_{k,l}(n) = A$ with $A$ a constant.  This solution is not  normalisable unless $A=0$, yielding a vanishing wave-function.  This is not different from the commutative free particle with zero energy, for which the wave-function is also just a constant and hence not normalisable unless the wave-function vanishes everywhere. 

The case $\varepsilon = 1$ is very similar to $\varepsilon = 0$.  In this case $\xi = -1$ and the difference equation  (\ref{eqn_hn_1}) becomes ($n\geq l+1$)
\begin{equation} \label{eqn_hn_E=1}
(n-l-1) \, h_{k,l}(n-2) + 2 n \, h_{k,l}(n-1) + (n+l+1) \, h_{k,l}(n) = 0.
\end{equation}
Set
\begin{equation}
h_{k,l}(n) = (-1)^n \, v_{k,l}(n)
\end{equation}
to obtain
\begin{equation} \label{eqn_vn_E=1}
(n-l-1) \, v_{k,l}(n-2) - 2 n \, v_{k,l}(n-1) + (n+l+1) \, v_{k,l}(n) = 0,
\end{equation}
which is the same as (\ref{eqn_hn_E=0}).  Hence a general solution is given by
\begin{equation}
h_{k,l}(n)  = (-1)^n \left[A + \frac{B}{(n-l+1)(n-l+2)...(n+l+1)}\right].
\end{equation}
The difference equation for $n=l+1$ becomes $h_{k,l}(l+1) = -h_{k,l}(l)$, which requires $B=0$ to yield $h_{k,l}(n) = (-1)^n A$ with $A$ a constant.  This solution is again not normalisable unless we set $A=0$ and the wave-function vanishes.   \\

Finally we consider $\varepsilon>1$.  The two values of $\mu$ are
\begin{equation}
-1 < \mu_1 = 1-2\varepsilon + 2\sqrt{\varepsilon (\varepsilon -1)}< 0, \quad \quad
\mu_2 = 1-2\varepsilon - 2\sqrt{\varepsilon (\varepsilon -1)} = \frac{1}{\mu_1} < -1.
\end{equation}
A general solution is given by (\ref{sol_hn_general}, \ref{sol_h_alpha_n}).
The asymptotic behaviours of the two solutions are
\begin{equation}
\left. h_{k,l}^{(\alpha)}(n)\right|_{n\rightarrow\infty} \approx \frac{\alpha_0 \mu_\alpha^n}{(n-l+1)(n-l+2)...(n+1)}, \quad \quad \alpha=1,2.
\end{equation}
Clearly the solution corresponding to $\mu_2<-1$ is not well-behaved in the asymptotic limit and not normalisable unless we require $A_2 = 0$ in (\ref{sol_hn_general}). The difference equation (\ref{eqn_hn_1}) at $n=l+1$ reads $h_{k,l}(l+1) = \xi h_{k,l}(l)$, which implies
\begin{equation} \label{condition_origin_E>1}
A_1 \alpha_0 \,\, \mu_1^l \sum_{s=0}^{l} \left(\frac{\mu_1}{2(\xi-\mu_1)}\right)^s \frac{(l+s)!}{(l+s+2)! \,\, (l-s)! \,\, s!} \left[\mu_1 - \xi (l+s+2)\right]=0.
\end{equation}
Noting that for $\varepsilon > 1$
\begin{eqnarray}
&\mu_1<0, \quad \quad \xi = 1-2\varepsilon<0, \quad \quad  \xi - \mu_1 = -2 \sqrt{\varepsilon (\varepsilon - 1)}<0,& \nonumber \\
&\mu_1 - \xi (l+s+2) = (l+s+1)(2\varepsilon - 1) + 2 \sqrt{\varepsilon (\varepsilon -1)} > 0, 
\end{eqnarray}
for non-negative $s$ and $l$, we conclude that each term in the LHS of (\ref{condition_origin_E>1}) is positive. The sum is therefore non-zero and $A_1 = 0$, yielding a trivial solution. \\

We therefore obtain the rather remarkable result that the free particle energy is bounded by $\varepsilon=1$ or $E = \frac{8\hbar^2}{\theta^2 m_0}$.
Note that the presence of the high energy cut-off for a free particle in 3D fuzzy space will certainly affect the spectrum of a particle in some external potential and modify the scattering data.  This result will clearly also have drastic physical implications at high energies (temperatures) and densities of Fermionic gases.
Such an upper bound on the energy of a particle also appears in DSR (Doubly/Deformed Special Relativistic) theories where a well-motivated presence of an invariant energy (generally regarded as the high energy cut-off of the theory) paves the way for deformations of usual relativity \cite{AmelinoCamelia:2000mn, Magueijo:2002am}.
These theories also include the discussion of modified dispersion relations and in particular
\cite{Chandra:2011nj} gives a robust treatment of the thermodynamics of an ideal gas in such a scenario.
These theories have also been related to certain kinds of non-commutative algebras \cite{KowalskiGlikman:2001ct}. \\

A more remarkable consequence is the existence of a complete duality between the high energy (ultra-violet) and low energy (infra-red) sectors on which we elaborate more below.  
As observed above the energy spectrum of the fuzzy free particle is bounded both from the below and above. The two boundaries correspond to the values $\varepsilon = 0$ and $\varepsilon = 1$ or $\varphi = 0$ and $\varphi = \pi$, respectively. Consider the transformation (we call it the energy reversal transformation)
\begin{equation}
\mathcal{E}: \varphi \rightarrow \pi - \varphi,
\end{equation}
which relates the two boundaries. This transformation actually relates a state with dimensionless energy parameter $\varepsilon$ to that with $1-\varepsilon$ and therefore turns the energy spectrum of the free particle upside down.  We note that
\begin{equation}
\mathcal{E}^2 = \mathds{1}.
\end{equation}
The eigenvalues of the operator $\mathcal{E}$ are $\pm 1$.  The action of the energy reversal operator on a state in the Hilbert space is given by
\begin{equation}
\mathcal{E}: \hat{\psi}_{\varphi,l,m} \rightarrow \hat{\psi}_{\pi - \varphi,l,m}.
\end{equation}
The eigenstates are simply found as:
\begin{eqnarray}
\mathcal{E} \hat{\psi}_{\pm} = \mathcal{E} \,\, \frac{1}{\sqrt{2}} \left[\hat{\psi}_{\varphi,l,m} \pm \hat{\psi}_{\pi - \varphi, l, m}\right] = \pm \frac{1}{\sqrt{2}} \left[\hat{\psi}_{\varphi,l,m} \pm \hat{\psi}_{\pi - \varphi, l, m}\right] = \pm  \hat{\psi}_{\pm}.
\end{eqnarray}
One can check that
\begin{equation}
h_{\pi - \varphi,l}^{(1)}(n) = \tan\left(\frac{\varphi}{2}\right) e^{in\pi} \, h_{\varphi,l}^{(1)*}(n),
\end{equation}
implying
\begin{equation}
\cos(\pi - \varphi) \, h_{\pi - \varphi,l}^{(1)}(l) - h_{\pi - \varphi,l}^{(1)}(l+1) = - \tan\left(\frac{\varphi}{2}\right) e^{il\pi} \left[\cos\varphi \, h_{\varphi,l}^{(1)*}(l) - h_{\varphi,l}^{(1)*}(l+1)\right],
\end{equation}
which in turn yields
\begin{equation}
e^{2i \delta_{\pi - \varphi,l}} =
\frac{\cos(\pi - \varphi) \, h_{\pi - \varphi,l}^{(1)}(l) - h_{\pi - \varphi,l}^{(1)}(l+1)}{\cos(\pi - \varphi) \, h_{\pi - \varphi,l}^{(1)*}(l) - h_{\pi - \varphi,l}^{(1)*}(l+1)} =
\frac{\cos\varphi \, h_{\varphi,l}^{(1)*}(l) - h_{\varphi,l}^{(1)*}(l+1)}{\cos\varphi \, h_{\varphi,l}^{(1)}(l) - h_{\varphi,l}^{(1)}(l+1)} = e^{-2i \delta_{\varphi,l}},
\end{equation}
and subsequently
\begin{equation} \label{delta_pi-phi}
\delta_{\pi - \varphi, l} = - \delta_{\varphi,l}.
\end{equation}
The standing wave solution (\ref{sol_hn_E<1_nonsingular}) for $\varphi \rightarrow \pi - \varphi$ gives
\begin{equation}
h_{\pi - \varphi,l}(n) = -(-1)^n \, \frac{A'}{A} \tan\left(\frac{\varphi}{2}\right) \, h_{\varphi,l}(n).
\end{equation}
Note that the normalisation constant $A$ depends on $\varphi$ and hence goes to $A'$. As $\hat{\mathcal{Y}}_{lm}$ remains invariant we have
\begin{equation} \label{psi_pi-phi}
\hat{\psi}_{\pi - \varphi,l,m}
= -(-1)^{\hat{N}} \, \frac{A'}{A} \tan\left(\frac{\varphi}{2}\right) \, \hat{\psi}_{\varphi,l,m},
\end{equation}
where $(-1)^{\hat{N}}$ is defined as:
\begin{equation}
(-1)^{\hat{N}} = \sum_{n=0}^{\infty} (-1)^n \hat{P}_{n/2}.
\end{equation}
Using (\ref{psi_pi-phi}) gives
\begin{equation}
\left(\hat{\psi}_{\pi - \varphi,l,m}, \hat{\psi}_{\pi - \varphi',l',m'}\right) 
= \left|\frac{A'}{A}\right|^2 \tan^2\left(\frac{\varphi}{2}\right) \, \left(\hat{\psi}_{\varphi,l,m}, \hat{\psi}_{\varphi',l',m'}\right).
\end{equation}
On the other hand  the orthogonality relation (\ref{orthogonality_psi}) implies
\begin{equation}
\left(\hat{\psi}_{\pi - \varphi,l,m}, \hat{\psi}_{\pi - \varphi',l',m'}\right) 
= \left(\hat{\psi}_{\varphi,l,m}, \hat{\psi}_{\varphi',l',m'}\right).
\end{equation}
Hence we must have
\begin{equation}
 \left|\frac{A'}{A}\right|^2 \tan^2\left(\frac{\varphi}{2}\right) = 1.
\end{equation}
This in turn gives the transformation of the wave functions under energy reversal as
\begin{equation} \label{symmetry_psi}
\mathcal{E}: \hat{\psi}_{\varphi,l,m} \rightarrow \hat{\psi}_{\pi - \varphi,l,m}  = e^{i\eta} \, (-1)^{\hat{N}} \, \hat{\psi}_{\varphi,l,m},
\end{equation}
$\eta$ being a real phase. From linearity this extends to any element $\hat\psi$ in the quantum Hilbert space ${\cal H}_q$  
\begin{equation}
\mathcal{E}: \hat{\psi} \rightarrow \hat{\mathcal{E}} \, \hat{\psi} = e^{i\eta} \, (-1)^{\hat{N}} \, \hat{\psi}.
\end{equation}
Let us calculate the commutator of $\hat{\mathcal{E}}$ with the Hamiltonian (\ref{Hamiltonian})
\begin{eqnarray}
[\hat{\mathcal{E}},\hat{H}_0] \hat{\psi}_{\varphi,l,m} &=& \hat{\mathcal{E}} \hat{H}_0 \hat{\psi}_{\varphi,l,m} - \hat{H}_0 \hat{\mathcal{E}}  \hat{\psi}_{\varphi,l,m}\nonumber \\
&=& \hat{\mathcal{E}} \, \frac{8\hbar^2 \varepsilon}{\theta^2 m_0}  \hat{\psi}_{\varphi,l,m} - \hat{H}_0 \hat{\psi}_{\pi - \varphi,l,m}  \label{E_H_comm_1}\nonumber \\
&=& \frac{8\hbar^2 \varepsilon}{\theta^2 m_0} \hat{\psi}_{\pi - \varphi,l,m} - \frac{8\hbar^2 (1-\varepsilon)}{\theta^2 m_0}  \hat{\psi}_{\pi - \varphi,l,m} \label{E_H_comm_2}\nonumber\\
&=&  \frac{8\hbar^2 (2\varepsilon-1)}{\theta^2 m_0}  \hat{\psi}_{\pi - \varphi,l,m}\nonumber \\
&=& \hat{\mathcal{E}} \left(2\hat{H}_0 -  \frac{8\hbar^2}{\theta^2 m_0}\right) \hat{\psi}_{\varphi,l,m}\nonumber \\
&=& \left(\frac{8\hbar^2}{\theta^2 m_0} - 2\hat{H}_0\right) \hat{\mathcal{E}} \hat{\psi}_{\varphi,l,m}.
\end{eqnarray}
Since the $\hat{\psi}_{\varphi,l,m}$ forms a complete orthonormal basis, this implies the operator relation:
\begin{equation}
[\hat{\mathcal{E}},\hat{H}_0] =  \hat{\mathcal{E}} \left(2\hat{H}_0 -  \frac{8\hbar^2}{\theta^2 m_0}\right) = \left(\frac{8\hbar^2}{\theta^2 m_0} - 2\hat{H}_0\right) \hat{\mathcal{E}},
\end{equation}
from which the anti-commutator can be read off:
\begin{equation}
\{\hat{\mathcal{E}}, \hat{H}_0\} = \frac{8\hbar^2}{\theta^2 m_0} \hat{\mathcal{E}}.
\end{equation}

\section{The 3D fuzzy well}
\label{five}
We first consider the piecewise constant potential of the form
\begin{equation}
\hat{V}(\hat{R}) =  \sum_{n=0}^{\infty} V(n)\hat{P}_{n/2}, \quad \quad
V(n) = \left\{
\begin{array}{cl}
V_{in} & {\rm for} \,\, 0\leq n \leq M \\
V_{out} & {\rm for} \,\, n \geq M+1.
\end{array}
\right.
\end{equation}
The Schr\"{o}dinger equation
\begin{equation} \label{Sch_eqn_well}
\hat{H} \hat{\psi} =  \left(\hat{H}_0 + \hat{V}(\hat{R})\right) \hat{\psi} = E \hat{\psi}
\end{equation}
has the solution of the form $\hat{\psi}_{k,l,m} = \hat{\mathcal{G}}_{k,l}(\hat{R}) \, \hat{\mathcal{Y}}_{lm}$.
Writing $\hat{\mathcal{G}}_{k,l}(\hat{R}) =  \sum_{n=0}^{\infty} \mathcal{G}_{k,l}(n)\hat{P}_{n/2}$ we get the equation for $\mathcal{G}_{k,l}(n)$ as
\begin{equation} \label{eqn_Gn_well}
(n-l-1) \, \mathcal{G}_{k,l}(n-2) - 2 n \,\, \zeta(n-1) \,\, \mathcal{G}_{k,l}(n-1) + (n+l+1) \, \mathcal{G}_{k,l}(n) = 0; \quad \quad n\geq l+1
\end{equation}
with
\begin{equation}
\zeta(n) = 1-\frac{\theta^2m_0}{4\hbar^2}\left(E-V(n)\right) = \left\{
\begin{array}{cl}
\zeta_{in} = \xi + \frac{\theta^2 m_0}{4\hbar^2}V_{in} & {\rm for} \,\, 0\leq n \leq M \\
\zeta_{out} = \xi + \frac{\theta^2 m_0}{4\hbar^2}V_{out} & {\rm for} \,\, n \geq M+1.
\end{array}
\right.
\end{equation}
Thus we get the following two equations for the two different regions:
\begin{eqnarray}
(n-l-1) \, \mathcal{G}_{k,l}(n-2) - 2 n \, \zeta_{in} \, \mathcal{G}_{k,l}(n-1) + (n+l+1) \, \mathcal{G}_{k,l}(n) = 0; & \quad \quad & l+1\leq n \leq M+1 \label{eqn_Gn_well_in} \\
(n-l-1) \, \mathcal{G}_{k,l}(n-2) - 2 n \, \zeta_{out} \, \mathcal{G}_{k,l}(n-1) + (n+l+1) \, \mathcal{G}_{k,l}(n) = 0; & \quad \quad & n \geq M+2.  \label{eqn_Gn_well_out}
\end{eqnarray}
The domains of $\mathcal{G}_{k,l}(n)$ in (\ref{eqn_Gn_well_in}) and (\ref{eqn_Gn_well_out}) are $n\in\{l-1,l,l+1,...,M+1\}$ and $n\in\{M,M+1,M+2,...\}$ respectively.
Thus if $\mathcal{G}_{k,l}^{(in)}(n)$ and $\mathcal{G}_{k,l}^{(out)}(n)$ are solutions of (\ref{eqn_Gn_well_in}) and (\ref{eqn_Gn_well_out}) respectively, the radial part of the wave function for the whole space is given by
\begin{equation}
\mathcal{G}_{k,l}(n) = \left\{
\begin{array}{cl}
\mathcal{G}_{k,l}^{(in)}(n) & {\rm for} \,\, l\leq n \leq M+1, \\
\mathcal{G}_{k,l}^{(out)}(n) & {\rm for} \,\, n\geq M
\end{array}
\right.
\end{equation}
with $\mathcal{G}_{k,l}^{(in)}(n)$ and $\mathcal{G}_{k,l}^{(out)}(n)$ matching at $n=M$ and at $n=M+1$, i.e.,
\begin{equation} \label{match}
\mathcal{G}_{k,l}^{(in)}(M) = \mathcal{G}_{k,l}^{(out)}(M), \quad \quad \mathcal{G}_{k,l}^{(in)}(M+1) = \mathcal{G}_{k,l}^{(out)}(M+1).
\end{equation}
Let us now consider a fuzzy spherical well with radius $R=\theta M/2$ and depth $V_0$. Here $M$ is an integer. 
This corresponds to
\begin{equation}
V_{in} = -V_0, \quad \quad V_{out} = 0, \quad \quad V_0>0
\end{equation}
which implies
\begin{equation}
\zeta_{in} = \xi - 2\upsilon_0 = 1-2\varepsilon_{in}, \quad \quad \zeta_{out} = \xi = 1-2\varepsilon, \quad \quad \upsilon_0 = \frac{\theta^2 m_0}{8\hbar^2}V_0, \quad \quad \varepsilon_{in} = \varepsilon+\upsilon_0.
\end{equation}
Since $-V_0$ is the minimum value of the potential,  $E \geq -V_0 \Rightarrow \varepsilon \geq -\upsilon_0$. \\

We start by assuming the $\upsilon_0-$value in the range $0<\upsilon_0<1$.
In this case the energy values satisfying $-1<-\upsilon_0<\varepsilon<0$ corresponds to $0<\varepsilon_{in}<\upsilon_0 <1$.
Thus the solution for $\mathcal{G}_{k,l}^{(in)} (n)$ is given by 
\begin{equation} \label{sol_Gn_in}
\mathcal{G}_{k,l}^{(in)}(n) = A_1 \left[\mathcal{G}_{k,l}^{(1in)}(n) - e^{2i\delta_{\varphi_{in},l}} \, \mathcal{G}_{k,l}^{(1in)*}(n)\right], \quad \quad n\geq l
\end{equation}
with
\begin{equation} \label{phi_in}
\varphi_{in} = 2 \sin^{-1} \left(\varepsilon_{in}^{1/2}\right), \quad \quad \delta_{\varphi_{in},l} = arg\left[\cos\varphi_{in} \,\, \mathcal{G}_{k,l}^{(1in)}(l) - \mathcal{G}_{k,l}^{(1in)}(l+1)\right]
\end{equation}
and
\begin{equation} \label{sol_Gn_1in}
\mathcal{G}_{k,l}^{(1in)}(n) = \frac{1}{2} \left(\frac{2}{\theta}\right)^{l} \csc \left(\frac{\varphi_{in}}{2}\right) e^{in\varphi_{in}} \sum_{s=0}^{l} \frac{e^{is\left(\varphi_{in} +\frac{\pi}{2}\right)}}{(2 \sin\varphi_{in})^s} \frac{(n-l)! \, (l+s)!}{(n+s+1)! \, (l-s)! \, s!}.
\end{equation}
A general solution to the equation for $\mathcal{G}_{k,l}^{(out)}(n)$, i.e., (\ref{eqn_Gn_well_out}) is given by
\begin{equation} \label{Gout}
\mathcal{G}_{k,l}^{(out)}(n) = B_1 \, \mathcal{G}_{k,l}^{(1out)}(n) + B_2 \, \mathcal{G}_{k,l}^{(2out)}(n)
\end{equation}
with
\begin{equation}
\mathcal{G}_{k,l}^{(\alpha \, out)}(n) = \mu_\alpha^n \sum_{s=0}^{l} \alpha_0 \left(\frac{\mu_\alpha}{2(\xi-\mu_\alpha)}\right)^s \frac{(n-l)! \,\, (l+s)!}{(n+s+1)! \,\, (l-s)! \,\, s!} \quad \quad ;\alpha=1,2
\end{equation}
the two values of $\mu_\alpha$ being given by
\begin{equation}
 \mu_1 = 1-2\varepsilon + 2\sqrt{-\varepsilon (1-\varepsilon)}>1, \quad \quad
0<\mu_2 = 1-2\varepsilon - 2\sqrt{-\varepsilon (1-\varepsilon)} = \frac{1}{\mu_1} < 1.
\end{equation}
$\mathcal{G}_{k,l}^{(1out)}(n)$ at $n\rightarrow\infty$ goes to infinity and hence we must have $B_1 =0$ and we get
\begin{equation} \label{Gout_final}
\mathcal{G}_{k,l}^{(out)}(n) = B_2 \, \mu_2^n \sum_{s=0}^{l} \alpha_0 \left(\frac{\mu_2}{2(\xi-\mu_2)}\right)^s \frac{(n-l)! \,\, (l+s)!}{(n+s+1)! \,\, (l-s)! \,\, s!}.
\end{equation}
The matching conditions (\ref{match}) quantizes the energy eigenvalues by the following equation
\begin{equation}
\frac{\mathcal{G}_{k,l}^{(1in)}(M)-e^{2i\delta_{\varphi_{in},l}}\mathcal{G}_{k,l}^{(1in)*}(M)}{\mathcal{G}_{k,l}^{(1in)}(M+1)-e^{2i\delta_{\varphi_{in},l}}\mathcal{G}_{k,l}^{(1in)*}(M+1)} = \frac{\mathcal{G}_{k,l}^{(2out)}(M)}{\mathcal{G}_{k,l}^{(2out)}(M+1)}.
\end{equation} 
These are the bound states and they vanish exponentially as $\frac{\mu_2^n}{n}$ with $\mu_2<1$ for $n\rightarrow\infty$.

Next the energy values satisfying $0<\varepsilon<1-\upsilon_0<1$ corresponds to $0<\upsilon_0<\varepsilon_{in}<1$.
The solution for $\mathcal{G}_{k,l}^{(in)} (n)$ is again given by (\ref{sol_Gn_in}), (\ref{phi_in}) and (\ref{sol_Gn_1in}).
On the other hand $\mathcal{G}_{k,l}^{(out)}(n)$ is given by
\begin{equation} \label{sol_Gn_out}
\mathcal{G}_{k,l}^{(out)}(n) = B_1 \, \mathcal{G}_{k,l}^{(1out)}(n) + B_2 \, \mathcal{G}_{k,l}^{(1out)*}(n)
\end{equation}
with
\begin{equation} \label{g1out}
\mathcal{G}_{k,l}^{(1out)}(n) = \frac{1}{2} \left(\frac{2}{\theta}\right)^{l} \csc \left(\frac{\varphi}{2}\right) e^{in\varphi} \sum_{s=0}^{l} \frac{e^{is\left(\varphi +\frac{\pi}{2}\right)}}{(2 \sin\varphi)^s} \frac{(n-l)! \, (l+s)!}{(n+s+1)! \, (l-s)! \, s!}
\end{equation}
and $\varphi = 2 \sin^{-1} \left(\varepsilon^{1/2}\right)$.
The matching condition (\ref{match}) gives
\begin{equation}
B_1 = \mathcal{B} A_1, \quad \quad B_2 = -\mathcal{B}^* e^{2i\delta_{\varphi_{in},l}} \, A_1
\end{equation}
with
\begin{eqnarray}
\mathcal{B} = &&
\left[ \mathcal{G}_{k,l}^{(1out)*}(M) \,\, \mathcal{G}_{k,l}^{(1in)}(M+1) - \mathcal{G}_{k,l}^{(1in)}(M) \,\, \mathcal{G}_{k,l}^{(1out)*}(M+1) \right. \nonumber \\
 && +\left. \left. e^{2i\delta_{\varphi_{in},l}} \left\{\mathcal{G}_{k,l}^{(1in)*}(M) \,\, \mathcal{G}_{k,l}^{(1out)*}(M+1) - \mathcal{G}_{k,l}^{(1out)*}(M) \,\, \mathcal{G}_{k,l}^{(1in)*}(M+1)\right\} \right]
\right/  \nonumber \\
&&
\left[\mathcal{G}_{k,l}^{(1out)*}(M) \,\, \mathcal{G}_{k,l}^{(1out)}(M+1) - \mathcal{G}_{k,l}^{(1out)}(M) \,\, \mathcal{G}_{k,l}^{(1out)*}(M+1)\right].
\end{eqnarray}
We can write $B_2 = -e^{2i\beta_l} B_1$ with $\beta_l = \delta_{\varphi_{in},l}-arg(\mathcal{B})$.
The asymptotic form of (\ref{g1out}) is given by
\begin{equation}
\mathcal{G}_{k,l}^{(1out)}(n)|_{n\rightarrow\infty} = \frac{1}{k}\left(\frac{2}{\theta}\right)^{l+1} \frac{e^{in\varphi}}{n^{l+1}}.
\end{equation}
Thus the wave function represents a scattering state describing an incident incoming spherical wave of energy $E$ getting fully reflected at the origin to produce an outgoing spherical wave of the same energy with a phase difference of $2\beta_l$.
If the potential is identically zero everywhere, one would get  $\hat{V}(\hat{R}) = 0 \Rightarrow \beta_l|_{\hat{V}(\hat{R})=0} = \delta_{\varphi,l}$.
Hence the phase shift is given by
\begin{equation}
\Delta_l = 2\left(\beta_l -\beta_l|_{\hat{V}(\hat{R})=0}\right) = 2\left(\delta_{\varphi_{in},l} - \delta_{\varphi,l} - arg(\mathcal{B})\right).
\end{equation}
The deviation in the phase shift from the commutative case will result in the corresponding deviation in the cross section of the scattering experiment.

Let us now consider the energy values $0<1-\upsilon_0<\varepsilon<1$ which corresponds to $\varepsilon_{in} >1$.
The solution for $\mathcal{G}_{k,l}^{(in)} (n)$ is given by 
\begin{equation} \label{Gin}
\mathcal{G}_{k,l}^{(in)}(n) = A_1 \, \mathcal{G}_{k,l}^{(1in)}(n) + A_2 \, \mathcal{G}_{k,l}^{(2in)}(n)
\end{equation}
with
\begin{equation}
\mathcal{G}_{k,l}^{(\alpha in)}(n) = \mu_{\alpha in}^n \sum_{s=0}^{l} \frac{1}{k} \left(\frac{2}{\theta}\right)^{l+1} \left(\frac{\mu_{\alpha in}}{2(\xi-2\upsilon_0-\mu_{\alpha in})}\right)^s \frac{(n-l)! (l+s)!}{(n+s+1)!(l-s)!s!} \quad \quad ;\alpha=1,2.
\end{equation}
The two values of $\mu_{\alpha in}$ are
\begin{equation}
-1 < \mu_{1in} = 1-2\varepsilon_{in} + 2\sqrt{\varepsilon_{in} (\varepsilon_{in} -1)}< 0, \quad \quad
\mu_{2in} = 1-2\varepsilon_{in} - 2\sqrt{\varepsilon_{in} (\varepsilon_{in} -1)} = \frac{1}{\mu_{1in}} < -1.
\end{equation}
The equation for $\mathcal{G}_{k,l}^{(in)}(n)$ for $n=l+1$ gives (see (\ref{eqn_Gn_well_in})) $\mathcal{G}_{k,l}^{(in)}(l+1) = \zeta_{in} \, \mathcal{G}_{k,l}^{(in)}(l)$ which relates $A_1$ and $A_2$ by $A_2 = -\mathcal{A}A_1$ with
\begin{equation}
\mathcal{A} = \frac{\mathcal{G}_{k,l}^{(1in)}(l+1) - \zeta_{in} \, \mathcal{G}_{k,l}^{(1in)}(l)}{\mathcal{G}_{k,l}^{(2in)}(l+1) - \zeta_{in} \, \mathcal{G}_{k,l}^{(2in)}(l)},
\end{equation}
and we get
\begin{equation} \label{Gin_1}
\mathcal{G}_{k,l}^{(in)}(n) = A_1 \left[\mathcal{G}_{k,l}^{(1in)}(n) - \mathcal{A} \, \mathcal{G}_{k,l}^{(2in)}(n)\right].
\end{equation}
On the other hand $\mathcal{G}_{k,l}^{(out)}(n)$ is given by (\ref{sol_Gn_out}) and (\ref{g1out}).
The matching condition (\ref{match}) gives
\begin{equation}
B_1 = \mathcal{B} A_1, \quad \quad B_2 = \mathcal{B}^* A_1
\end{equation}
with
\begin{eqnarray}
\mathcal{B} = &&
\left[ \mathcal{G}_{k,l}^{(1in)}(M) \,\, \mathcal{G}_{k,l}^{(1out)*}(M+1) - \mathcal{G}_{k,l}^{(1out)*}(M) \,\, \mathcal{G}_{k,l}^{(1in)}(M+1) \right. \nonumber \\
&& +\left. \left. \mathcal{A} \left\{\mathcal{G}_{k,l}^{(1out)*}(M) \,\, \mathcal{G}_{k,l}^{(2in)}(M+1) - \mathcal{G}_{k,l}^{(2in)}(M) \,\, \mathcal{G}_{k,l}^{(1out)*}(M+1)\right\} \right] \right/
\nonumber \\
&& \left[\mathcal{G}_{k,l}^{(1out)}(M) \,\, \mathcal{G}_{k,l}^{(1out)*}(M+1) - \mathcal{G}_{k,l}^{(1out)*}(M) \,\,  \mathcal{G}_{k,l}^{(1out)}(M+1)\right].
\end{eqnarray}
Here we have used the fact that $\mathcal{G}_{1in}(n)$, $\mathcal{G}_{2in}(n)$ and $\mathcal{A}$ are real.
Writing $B_2 = -e^{2i\beta_l} B_1$ with $\beta_l = \frac{\pi}{2} - arg(\mathcal{B})$ gives the phase shift as
\begin{equation}
\Delta_l = 2\left(\beta_l -\beta_l|_{\hat{V}(\hat{R})=0}\right) = \pi - 2\left(\delta_{\varphi,l} + arg(\mathcal{B})\right).
\end{equation}

\begin{figure}[h]
\begin{center}
\scalebox{0.60}{\includegraphics{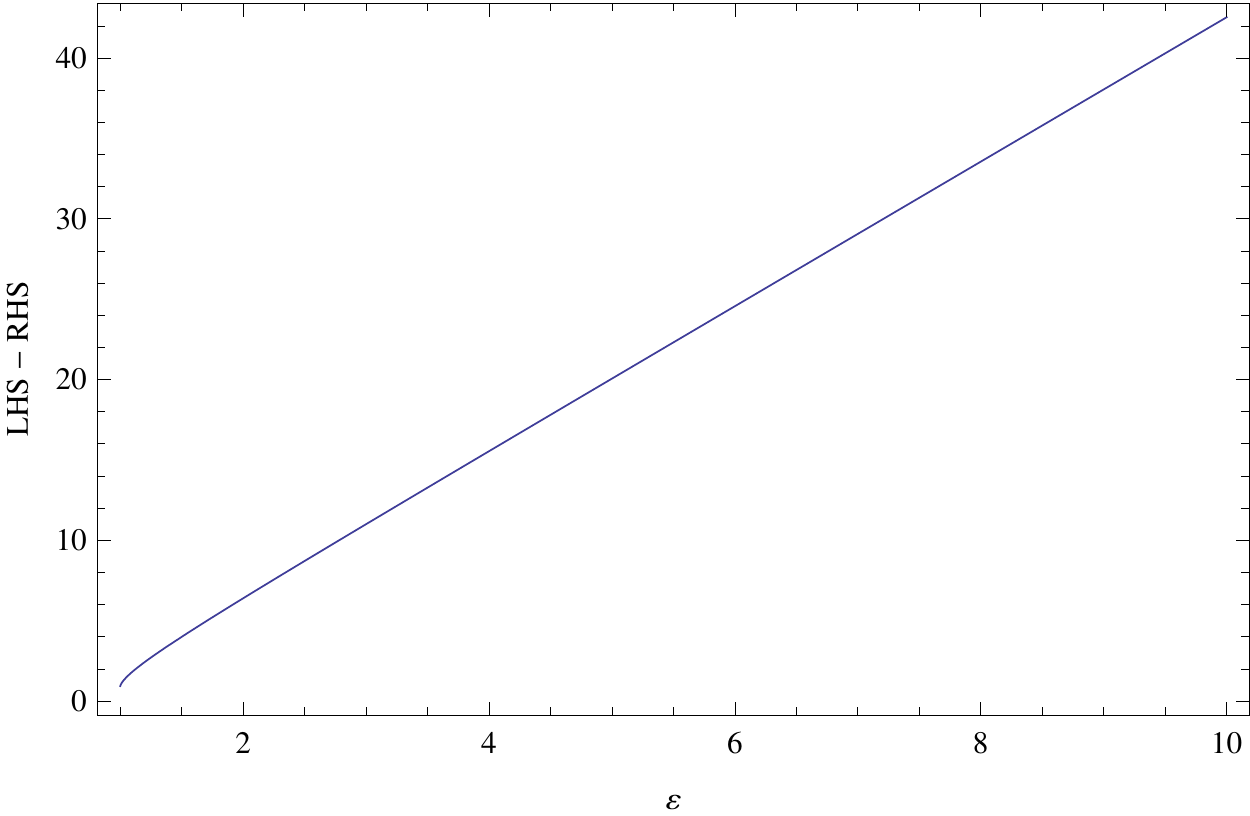}}
\scalebox{0.60}{\includegraphics{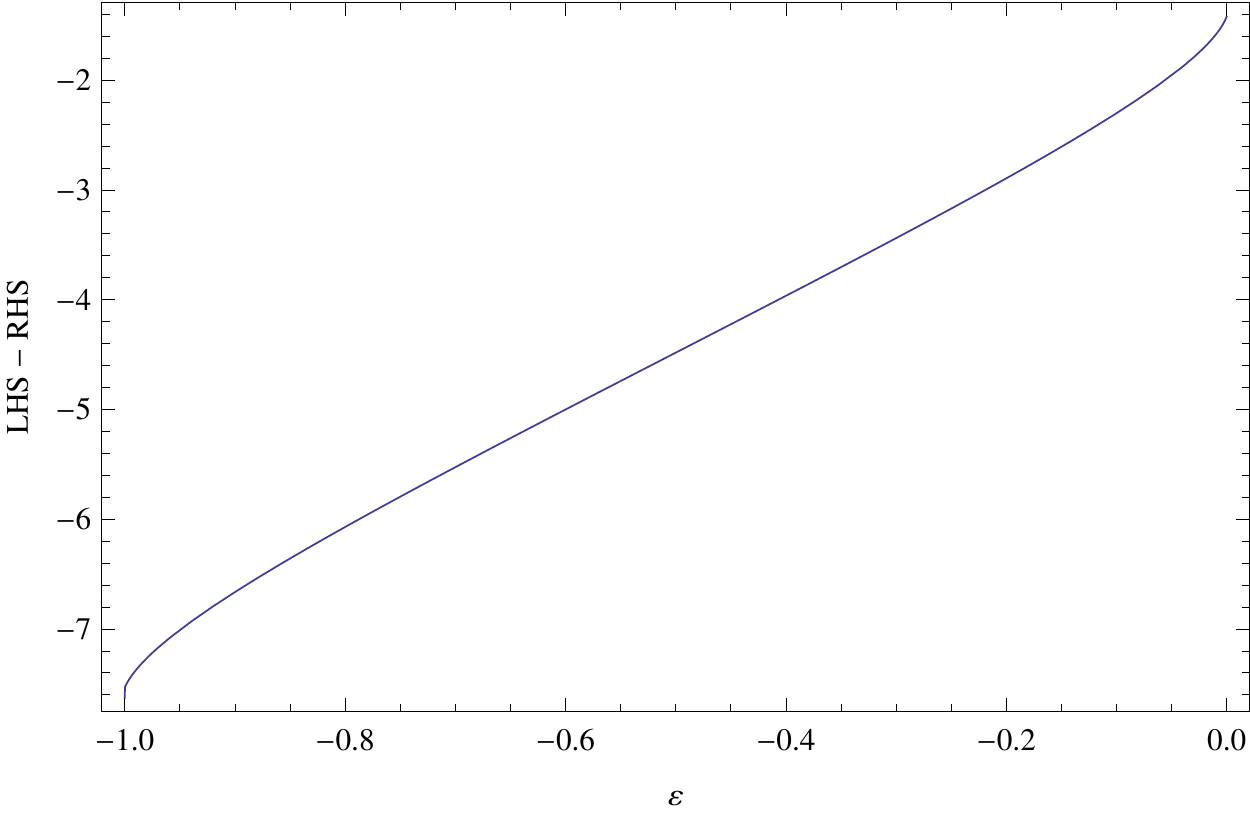}}
\end{center}
\caption{\label{fig_match_exotic} Typical plots of the difference between the LHS and the RHS of the matching conditions (\ref{match_exotic}) and (\ref{match_midrange}) with the dimensionless energy $\varepsilon$. The left one corresponds to the case $\upsilon_0 <1$ while the right one corresponds to $\upsilon_0 >1$. Various parameters for the plots shown are $M=100; \, \upsilon_0 = 0.5 \, {\rm (left)}, 2 \, {\rm (right)}; \, l=10$. Note that the effect of non-commutative parameter  $\theta$ is hidden inside the dimensionless potential depth $\upsilon_0$.}
\end{figure}

Finally the energy values $\varepsilon>1$ corresponds to $\varepsilon_{in}>1$.
The solution for $\mathcal{G}_{k,l}^{(in)} (n)$ is given by equations (\ref{Gin} - \ref{Gin_1}).
On the other hand the solution for $\mathcal{G}_{k,l}^{(out)}(n)$ is given by
\begin{equation}
\mathcal{G}_{k,l}^{(out)}(n) = B_1 \, \mu_{1}^n \sum_{s=0}^{l} \frac{1}{k} \left(\frac{2}{\theta}\right)^{l+1} \left(\frac{\mu_{1}}{2(\xi-\mu_{1})}\right)^s \frac{(n-l)! (l+s)!}{(n+s+1)!(l-s)!s!}
\end{equation}
with
\begin{equation}
-1 < \mu_1 = 1-2\varepsilon + 2\sqrt{\varepsilon (\varepsilon -1)}< 0.
\end{equation}
The matching conditions (\ref{match}) quantizes the energy eigenvalues by the following equation
\begin{equation} \label{match_exotic}
\frac{\mathcal{G}_{k,l}^{(1in)}(M)-\mathcal{A} \, \mathcal{G}_{k,l}^{(2in)}(M)}{\mathcal{G}_{k,l}^{(1in)}(M+1)-\mathcal{A} \, \mathcal{G}_{k,l}^{(2in)}(M+1)} = \frac{\mathcal{G}_{k,l}^{(out)}(M)}{\mathcal{G}_{k,l}^{(out)}(M+1)}.
\end{equation}
These unusual bound states at ultra-high energy values seems quite counter-intuitive and their existence may have certain unphysical implications.
In fact numerical evidences suggest that there is no real energy value for which the matching condition (\ref{match_exotic}) is satisfied.
The difference between the LHS and the RHS of  (\ref{match_exotic}) remains positive in the region $\varepsilon > 1$ and it increases monotonically with the increase in $\varepsilon$.
It never equals zero and hence there does not exist any bound state for $\varepsilon > 1$ (see figure \ref{fig_match_exotic} for a typical plot of this difference with the dimensionless energy $\varepsilon$).
We also note that there is an upper bound on the energy eigenvalues of the scattering states, i.e., $\varepsilon = 1$ beyond which there is no state possible.
This is a direct result of the presence of the high energy cut-off in the spectrum of the free particle found in section \ref{four}.

Let us now consider the other case when $\upsilon_0>1$. For this choice of $\upsilon_0$ we get the following solutions:
\begin{enumerate}
\item \underline{For $-\upsilon_0 < \varepsilon < -\upsilon_0 + 1 < 0$}, we have $0 < \varepsilon_{in} < 1$. 
The analysis follows that of the case of $0<\upsilon_0<1$ and we get quantized bound states.
\item \underline{For $-\upsilon_0+1 < \varepsilon < 0$}, we have $\varepsilon_{in} > 1$.
The solution for $\mathcal{G}_{k,l}^{(in)} (n)$ is given by equations (\ref{Gin} - \ref{Gin_1}).
The solution for $\mathcal{G}_{k,l}^{(out)} (n)$ is given by equations (\ref{Gout} - \ref{Gout_final}). 
The matching condition quantizes these bound states by the following condition:
\begin{equation} \label{match_midrange}
\frac{\mathcal{G}_{k,l}^{(1in)} (M) - \mathcal{A} \, \mathcal{G}_{k,l}^{(2in)} (M)}{\mathcal{G}_{k,l}^{(1in)} (M+1) - \mathcal{A} \, \mathcal{G}_{k,l}^{(2in)} (M+1)} = \frac{\mathcal{G}_{k,l}^{(out)} (M)}{\mathcal{G}_{k,l}^{(out)} (M+1)}.
\end{equation}
As in the case of the equation (\ref{match_exotic}) numerical evidence suggests that there exists no real energy value for which the above condition is satisfied (see figure 1) and hence these are not bound states as one may naively expect.  Indeed,  the spectrum is simply empty in this energy region.
\item \underline{For $0 < \varepsilon < 1$}, we have $\varepsilon_{in} > 1$.
The analysis again follows that of the case of $0<\upsilon_0<1$ and we get scattering states with a phase shift.
\item \underline{For $\varepsilon > 1$}, we have $\varepsilon_{in} > 1$.
Again the analysis follows that of the case of $0<\upsilon_0<1$ and the numerical evidences suggest no presence of any state in this regime.
\end{enumerate}

\section{An alternative approach: description in terms of Hypergeometric functions} 
\label{six}

This section gives a useful complimentary solution to the problem of the 3D fuzzy well using the technique invented by G\'alikov\'a et al \cite{press, Galikova:2011zf}.
This alternative approach gives the free particle wave functions in terms of well-known Hypergeometric functions, which turns out to be useful for solving the infinite well as demonstrated below.

First we introduce some notations.
Consider a function $g(n)$ defined on the non-negative integers via the series expansion $g(n)=\sum_{k=0}^{\infty} c_k n^k$. Associated with $g(n)$ is the function $\bar{g}(n)=\sum_{k}^\infty c_k n!/(n-k)!$ which satisfies
\begin{equation}
	g(n)=e^{-n}\sum_{k=0}^\infty \bar{g}(k)\frac{n^k}{k!}.
	\label{gandgbar}
\end{equation}
The motivation for this definition of $\bar{g}(n)$ is that, on the operator level, $\bar{g}(\hat{N})=\no{g(\hat{N})}$ where $\no{\ }$ denotes normal ordering of each power of $\hat{N}$ with respect to the bosonic vacuum of the Schwinger representation. This follows from the fact that $\no{\hat{N}^k}=\hat{N}!/(\hat{N}-k)!$, which can be easily proved using induction. In what follows we will abuse notation and simply write $\bar{g}(n)=\no{g(n)}$, understanding this to imply the operator relation above.\\

As before we will factorise the state of interest into an angular and radial part. The latter is a function of $\hat{N}$ and may be represented as $\bar{g}(\hat{N})=\no{g(\hat{N})}$. Following \cite{Galikova:2011zf} we write
\begin{equation}	\psih_{lm}=\sum_{(lm)}\frac{(\hat{a}^\dagger_1)^{m_1}(\hat{a}^\dagger_2)^{m_2}}{m_1!\,m_2!}\bar{g}(\hat{N})\frac{(\hat{a}_1)^{n_1}(-\hat{a}_2)^{n_2}}{n_1!\,n_2!}
\label{eq:standard-eigenfunctions}
\end{equation}
where the summation is restricted to non-negative integers $(m_1,m_2,n_1,n_2)$ satisfying $m_1+m_2=n_1+n_2=l$ and $m_1-m_2-n_1+n_2=2m$. As before $\psih_{lm}$ is an eigenstate of $\hat{L}^2$ and $\hat{L}_3$. However, due to its placement in the expression, the radial function $\bar{g}(\hat{N})$ is now related to the original $\hat{h}_{kl}(\hat{N})$ in \eqref{psi_hR_ylm} via a shift in its argument. We return to this point shortly. It is clear that the functions $g(n)$ and $\bar{g}(n)$ encode the same information and may be used interchangeably to describe the radial wave function. Our strategy will be to convert the fuzzy Schr\"{o}dinger equation, which is fundamentally a difference equation for $\bar{g}(n)$, into a differential equation for $g(n)$ which is solvable using standard techniques. We then transform back to $\bar{g}(n)$ and impose the appropriate boundary conditions. This step is necessary since in the $g(n)$ description these conditions are non-local in nature. To determine the action of the Laplacian
\begin{equation}
	\hat{\Delta}_\theta\psih_{lm}=-\frac{2}{\theta(\hat{R}+\tfrac{\theta}{2})}[\hat{a}^\dagger_\alpha,[\hat{a}_\alpha,\psih_{lm}]]
\label{eq:nc-laplacian}
\end{equation}
on $g(\hat{N})$ we take note of two results from \cite{Galikova:2011zf}, namely:
\begin{equation}
		-[\hat{a}^\dagger_\alpha,[\hat{a}_\alpha,\psih_{lm}]] =\sum_{(lm)}\frac{(\hat{a}^\dagger_1)^{m_1}(\hat{a}^\dagger_2)^{m_2}}{m_1!\,m_2!}\no{[\hat{N}\,g''(\hat{N})+2(l+1)\,g'(\hat{N})]}\frac{(\hat{a}_1)^{n_1}(-\hat{a}_2)^{n_2}}{n_1!\,n_2!}
	\label{eq:nc-laplacian-action}
\end{equation}
and
\begin{equation}
	\hat{N}\psih_{lm} =\sum_{(lm)}\frac{(\hat{a}^\dagger_1)^{m_1}(\hat{a}^\dagger_2)^{m_2}}{m_1!\,m_2!} \no{\left[(\hat{N}+l)\,g(\hat{N})+\hat{N}\,g'(\hat{N})\right]} \frac{(\hat{a}_1)^{n_1}(-\hat{a}_2)^{n_2}}{n_1!\,n_2!}.
\label{eq:r-times-wave}
\end{equation}
Using these identities the free particle Sch{\"o}dinger equation $-\frac{\hbar^2}{2m_0}\Delta_\theta\psih_{lm}=E\psih_{lm}$ can be reduced to
\begin{equation}
	\no{\left[\hat{N}\,g''(\hat{N}) + [2(l+1)+\kappa^2\hat{N}]\,g'(\hat{N}) + \kappa^2[\hat{N}+l+1]\,g(\hat{N})\right]}\ = 0
\label{eq:free-sol3}
\end{equation}
where $\kappa=\theta k/2$ and $E=\hbar^2 k^2/(2m_0)$. This result may be read as a differential equation for $g(n)$ or can be converted into a difference equation for $\bar{g}(n)$ by performing the normal ordering. The latter is easily accomplished as follows. Consider, for example, the $\hat{N}g'(\hat{N})$ term. From \eqref{gandgbar} we see that $n g'(n)=\exp[-n]\sum_{k=0}^\infty k[\bar{g}(k)-\bar{g}(k-1)] n^k/k!$ which translates into $\hat{N}[\bar{g}(\hat{N})-\bar{g}(\hat{N}-1)]=\no{\hat{N}g'(\hat{N})}$ on the operator level. \emph{Here, by definition, $\bar{g}(-1)=0$.} Treating the other terms in a similar way produces
\begin{equation}
(n+2(l+1))\bar{g}(n+1) + (\kappa^2 -2) (n+l+1) \bar{g}(n) + n \bar{g}(n-1) = 0 \quad \quad \quad n\geq0,
\label{eq:recursion-final}
\end{equation}
which matches \eqref{eqn_hn} upon identifying $\bar{g}(n)$ with $h_{kl}(n+l)$. Our goal is to solve for $\bar{g}(n)$ by using the solutions of the associated differential equation
\begin{equation}
		n g''(n) + (2(l+1)+\kappa^2n)\,g'(n) + \kappa^2(n+l+1)\,g(n)=0\quad\quad\quad n\geq0.
\label{eq:de-final}
\end{equation}
To better understand how the solutions of \eqref{eq:recursion-final} and \eqref{eq:de-final} are related we write $g(n)$ as $g(n)=\exp[-n]\sum_{k=-\infty}^\infty p_k n^k$ where negative powers of $n$ are now included. Direct substitution into \eqref{eq:de-final} reveals that the $p_n$ coefficients satisfy
\begin{equation}
(n+2(l+1))(1+n)p_{n+1} + (\kappa^2 -2) (n+l+1)p_n + p_{n-1} = 0 \quad\quad\quad n\in\mathbb{Z},
\label{eq:recursion-p}
\end{equation}
which indeed reduces to \eqref{eq:recursion-final} for $n>0$ when we set $p_n=\bar{g}(n)/n!$. However, for \eqref{eq:recursion-final} and \eqref{eq:recursion-p} to agree when $n=0$ requires that $p_{-1}=0$ since $\bar{g}(-1)=0$ by definition. In fact, $p_{-1}=0$ implies that $p_n=0$ for all $n<0$ as well. The conclusion is therefore as follows: \emph{Solutions of the differential equation \eqref{eq:de-final} which are non-singular at $n=0$ provide, through $\{\bar{g}(n)=p_n n!\ :\ n\geq0\}$, a complete solution to the difference equation in \eqref{eq:recursion-final} for all $n\geq 0$. However, singular solutions of \eqref{eq:de-final} only produce a solution of \eqref{eq:recursion-final} for $n>0$.} To deal with singular solutions efficiently in what follows it is useful to extend the normal ordering procedure to negative powers of $n$. Since $n^{-m}=\exp[-n]\sum_{k=-m}^\infty n^k/(k+m)!$ and we are interested only in $p_n$ for $n\geq 0$, it is therefore natural to define $\no{n^{-m}}=n!/(n+m)!$.\\

Returning to the differential equation for $g(n)$ we find two linearly independent solutions
\begin{align}
		g_J(n)&=n^{-(l+1/2)}e^{-\kappa^2n/2} J_{l+\frac{1}{2}}\!\left( \kappa n \sqrt{1-\kappa^2/4}\right)\quad\ {\rm and}\ \quad	g_Y(n)&=n^{-(l+1/2)}e^{-\kappa^2n/2} Y_{l+\frac{1}{2}}\!\left( \kappa n \sqrt{1-\kappa^2/4}\right).\nonumber
\end{align}
Here $J_\nu(x)$ and $Y_\nu(x)$ are Bessel functions of the first and second kind respectively. The latter is known to be singular at the origin. To find $\bar{g}_J(n)=\no{g_J(n)}$ and $\bar{g}_Y(n)=\no{g_Y(n)}$ we use
\begin{equation}
	J_\nu(x)=(x/2)^\nu\sum_{k=0}^\infty \frac{(-x^2/4)^k}{k!\,\Gamma(\nu+k+1)}\quad\ \ {\rm and}\ \ \quad Y_{l+1/2}=(-1)^{l+1} J_{-(l+1/2)},
\end{equation}
together with the identity $\no{\exp[-an]n^k}\,=(1-a)^{n-k}\no{n^k}$ from \cite{Galikova:2011zf} to obtain
\begin{align}
	\bar{g}_J(n)&=\left(1-\frac{\kappa^2}{2}\right)^{n+l'}\left(\frac{\gamma}{2}\right)^{l'}\frac{1}{\Gamma(l'+1)}\,{_2F_1}\left[\frac{1-n}{2},-\frac{n}{2};l'+1;-\gamma^2\right],\\
	\bar{g}_Y(n)&=(-1)^{l+1}\left(1-\frac{\kappa^2}{2}\right)^{n+l'}\left(\frac{\gamma}{2}\right)^{-l'} \frac{\Gamma(1+n)}{\Gamma(1-l')\Gamma(2+2 l+n)}\,{_2F_1}\left[-l'-\frac{n}{2},-l-\frac{n}{2};1-l';-\gamma^2\right],
\end{align}
where $l'=l+1/2$, $\gamma=\kappa\sqrt{1-\kappa^2/4}/(1-\kappa^2/2)$ and ${_2F_1}$ is the hypergeometric function \cite{abromowitz}. In the absence of a potential $\bar{g}_J(n)$ is therefore the global solution to the radial free particle equation \eqref{eq:recursion-final}. It can be simplified further using identities (15.3.21) and (15.4.6) in \cite{abromowitz} to obtain
\begin{equation}
	\bar{g}_J(n)\propto\frac{n!}{\Gamma(n+l+3/2)}\ P_n^{(l+1/2,l+1/2)}\left(1-\frac{\kappa^2}{2}\right),
\label{eq:nbj-to-poly2}
\end{equation}
where $P_n^{(\alpha,\beta)}$ is the Jacobi polynomial.
This result will be useful in what follows, and also allows us to easily verify a number of earlier results. 
For example, consider the asymptotic behaviour of $P_n^{(\alpha,\beta)}(x)$ as $n\rightarrow\infty$.
It is known from theorems 8.21.7 and 8.21.8 in \cite{szego} that unless $x\in[-1,1]$, $P_n^{(\alpha,\beta)}(x)$ will diverge exponentially, thereby rendering $\bar{g}_J(n)$ of (\ref{eq:nbj-to-poly2}) inadmissible as a wave function.
This places an upper bound of two on $\kappa$, which translates into an energy bound of $E<8\hbar^2/(\theta^2 m_0)$; in agreement with the discussion above. 
Furthermore, by applying identity (22.15.1) in \cite{abromowitz} one can easily verify that in the commutative limit $\bar{g}_J$ reduces to the standard expression in terms of a spherical Bessel function.\\

When dealing with piecewise constant potentials we typically need to solve the same free particle equation as in \eqref{eq:recursion-final} but with the energy $E$ shifted by a constant (say $V$). For bound states with $E<V$ we define
\begin{equation}
	\kappa'=\frac{\theta}{2}\sqrt{\frac{2m_0(V-E)}{\hbar^2}}
	\label{kappaprimedef}
\end{equation}
in which case the general local solution to \eqref{eq:recursion-final} is a linear combination of $\bar{g}_J(n)$ and $\bar{g}_Y(n)$ with $\kappa$ replaced by $i\kappa'$. Both these solutions diverge as $n\rightarrow\infty$, but the combination $\bar{g}_H(n)=\bar{g}_J(n)+i\bar{g}_Y(n)$ is exponentially decaying. It is the analogue of the Hankel function of the first kind. The two terms in $\bar{g}_H(n)$ be combined into a single hypergeometric function by first applying (15.3.21) from \cite{abromowitz} to each term and then using (15.3.9) to combine them. This produces
\begin{equation}
	\bar{g}_H(n) = C\frac{(4+\kappa'^2)^{-n}\Gamma(1+n)}{\Gamma(2+l+n)}\ {_2F_1}\left(\frac{3}{2}+l+n\,,2+2l+n\,;3+2l+2n\,;\frac{4}{4+\kappa'^2}\right)
\label{eq:hankel-single}
\end{equation}
with $C$ a constant. \\
 
Finally, let us consider the problem of the fuzzy spherical potential well of radius $\frac{\theta M}{2}$.
As already pointed out the Schr\"{o}dinger equation (\ref{Sch_eqn_well}) can be recast as a difference equation as was done to obtain \eqref{eq:recursion-final}.
If we take the potential to be equal to zero inside the well and V outside, the resulting equations for the bound states ($E<V$) are
\begin{align}
\left[n+2(l+1)\right]\bar{g}(n+1) + (\kappa^2 -2) (n+l+1) \bar{g}(n) + n \bar{g}(n-1) &= 0 \ \ \quad \quad 0\leq n \leq \mathcal{M}, \label{eq:recursion-well_1}\\
\left[n+2(l+1)\right]\bar{g}(n+1) + (-\kappa'^2 -2) (n+l+1) \bar{g}(n) + n \bar{g}(n-1) &= 0 \ \ \quad \quad n>\mathcal{M}
\label{eq:recursion-well}
\end{align}
with $\kappa'$ as in \eqref{kappaprimedef} and $\mathcal{M}=M-l$. We note that $\mathcal{M}$ acts as the effective radius of the well in the expressions above. The shift $M\rightarrow M-l$ occurs when the potential operator $\hat{V}(\hat{R})$ is moved past the $(\hat{a}_1^\dag)^{m_1}(\hat{a}_2^\dag)^{m_2}$ factors appearing in $\hat{\psi}_{lm}$ in order to act on $\bar{g}(\hat{N})$. 
The above set of equations are same as the ones given in (\ref{eqn_Gn_well_in}) and (\ref{eqn_Gn_well_out}) with $\bar{g}(n) = \mathcal{G}_{k,l}(n+l)$.
Based on the results obtained earlier we can immediately conclude that
\begin{align}
	\bar{g}(n)&=\alpha_1 \bar{g}_J(n)\quad {\rm for} \quad n=0,\ldots,\mathcal{M},\mathcal{M}+1\\
	{\rm and}\ \ \bar{g}(n)&=\alpha_2 \bar{g}_H(n)\quad {\rm for} \quad n=\mathcal{M},\mathcal{M}+1,\ldots
\end{align}
with $\alpha_{1,2}$ normalisation constants. The two solutions overlap at $n=\mathcal{M},\mathcal{M}+1$ which implies the matching condition
\begin{equation}
	\frac{\bar{g}_J(\mathcal{M}+1)}{\bar{g}_J(\mathcal{M})}=\frac{\bar{g}_H(\mathcal{M}+1)}{\bar{g}_H(\mathcal{M})}.
	\label{eq:well-matching2}
\end{equation}
This constraint determines the possible values of the energy $E$ which enters through \mbox{$\kappa=\theta\sqrt{2m_0E}/(2\hbar)$} on the left and $\kappa'=\theta\sqrt{2m_0(V-E)}/(2\hbar)$ on the right. In the case of an infinitely deep well the form of $\bar{g}_H$ given in \eqref{eq:hankel-single} can be used to determine the behaviour of the right hand side of \eqref{eq:well-matching2} as $V,\kappa'\rightarrow\infty$. We find that
\begin{equation}
	\frac{\bar{g}_H(\mathcal{M}+1)}{\bar{g}_H(\mathcal{M})}=\frac{1-\mathcal{M}}{(\mathcal{M}+l+2)\kappa'^2}+\mathcal{O}(\kappa'^{-4}).
\label{eq:hankel-expand}
\end{equation}
In this limit the right hand side of \eqref{eq:well-matching2} therefore vanishes, and so too must $\bar{g}_J(\mathcal{M}+1)$. Using \eqref{eq:nbj-to-poly2} this condition can be expressed as
\begin{equation}
	P_{M-l+1}^{(l+1/2,l+1/2)}\left(1-\frac{\kappa^2}{2}\right)=0
\end{equation}
 and so the spectrum is determined by the zeros of the Jacobi polynomials. In stark contrast to the commutative case each angular momentum sector contains only a finite number of $M-l+1$ bound states. Furthermore, there are no bound states with an angular momentum greater than $M=2R/\theta$.
Note that for an infinitely deep well the wave function outside the well vanishes and the wave function inside satisfies (\ref{eq:recursion-well_1}). Existence of a non-trivial solution to such an equation requires $\mathcal{M} \geq 0 \Rightarrow l \leq M$.
The combination of non-commutativity and a finite system size therefore introduces cut-offs in both energy and angular momentum and renders the quantum mechanical state space finite dimensional.

\section{Summary and conclusions}

The aim of this paper was to establish a formalism for quantum mechanics on fuzzy 3D spaces and to solve the Schr{\"o}dinger equation for the free particle, 3D fuzzy finite and infinite wells.
As a consistency check it was verified that all the results had the appropriate commutative limits.
The most important result was that the spectra of the fuzzy free particle exhibit a high energy cut-off.
This also results in the presence of an upper bound on the possible energy eigenvalues
for the fuzzy potential well of finite depth.
In addition the free particle spectrum showed a remarkable duality between the high and low energy sectors.
The dispersion relation of the free particle got modified in order to accommodate the presence of the high energy cut-off.
The phase difference between the incoming and the outgoing spherical waves in the free particle wave functions get an extra energy dependence unlike the commutative case where they depend only on the angular momentum of the waves.
The phase shifts for the scattering around a 3D fuzzy spherical potential well have been calculated.
For the well of (dimensionless) potential depth $\upsilon_0 >1$ there exists a range of energy values sandwiched between the bound states at the bottom and the scattering states at the top of the spectrum where there does not exist any state at all.
In case of infinite potential well, the finiteness of the system size introduces an upper cut-off (given by the radius of the system) for the angular momentum values.
The next step is to investigate further the physical consequences of these results and in particular the effect of the high energy cut-off.
Clearly, the latter must have profound consequences for the thermodynamics of Fermi gases at high densities and temperatures.
These issues will be the theme of a forthcoming paper. 

\appendix
\section{Some useful results}
It is straightforward to prove the following result by mathematical induction for a non-negative integer $n$
\begin{equation}\label{a_alpha_rn_com}
 \left[a_\alpha,\hat{R}^n\right]=a_\alpha\left(\hat{R}^n-\left(\hat{R}-\frac{\theta}{2}\right)^n\right).
\end{equation}
Thus for any Taylor series expandable function $\hat{g}(\hat{R})$ we get 
\begin{equation} \label{a_alpha_fr_com}
\left[a_{\alpha},\hat{g}(\hat{R})\right] = a_{\alpha}\left(\hat{g}(\hat{R})-\hat{g}\left(\hat{R}-\frac{\theta}{2}\right)\right).
\end{equation}
In essentially similar manner we can also prove
\begin{equation} \label{a_alpha_dagger_fr_com}
\left[ a_{\alpha}^{\dagger},\hat{g}(\hat{R}) \right] = \left(\hat{g}\left(\hat{R}-\frac{\theta}{2}\right)-\hat{g}(\hat{R})\right)a_{\alpha}^{\dagger}.
\end{equation}
$\hat{L}_{\pm} = \hat{L}_1 \pm i\hat{L}_2$ given by (\ref{l-}) satisfies Leibnitz rule (see (\ref{li_leibnitz})) which gives
\begin{equation}
 \left(\hat{L}_-\hat{\phi}\right)a_\alpha=\hat{L}_-\left(\hat{\phi} a_\alpha\right)+\hbar \delta_{\alpha 2}\hat{\phi}a_1, \quad \quad
  a_\alpha\left(\hat{L}_-\hat{\phi}\right)=\hat{L}_-\left(a_\alpha\hat{\phi}\right)+\hbar \delta_{\alpha 2}a_1\hat{\phi}.
\end{equation}
Using the Leibnitz rule again and again we get
\begin{equation}
 \left(\hat{L}_-^n\hat{\phi}\right)a_\alpha=\hat{L}_-^n\left(\hat{\phi} a_\alpha\right)+n\hbar \delta_{\alpha 2}\left(\hat{L}_-^{n-1}\hat{\phi}\right)a_1, \quad \quad
  a_\alpha\left(\hat{L}_-^n\hat{\phi}\right)=\hat{L}_-^n\left(a_\alpha\hat{\phi}\right)+n\hbar \delta_{\alpha 2}a_1\left(\hat{L}_-^{n-1}\hat{\phi}\right).
\end{equation}
Now, using the definition of $\hat{\mathcal{Y}}_{lm}$, i.e., equation (\ref{ylm}) and the above results we get
\begin{equation}
 \left[a_\alpha,\hat{\mathcal{Y}}_{lm}\right]=c_{lm}\hat{L}_-^{l-m}\left[a_\alpha,\hat{\mathcal{Y}}_{ll}\right]+(l-m)\hbar\delta_{\alpha 2}\frac{c_{lm}}{c_{l,m+1}}\left[a_1,\hat{\mathcal{Y}}_{l,m+1}\right].
\end{equation}
 In particular, for $m=l$
\begin{equation}
 \left[a_{\alpha},\hat{\mathcal{Y}}_{ll}\right]=l\delta_{\alpha 1}\left(a_1^{\dagger}\right)^{l-1}\left(a_2\right)^l,
\end{equation}
and we get
\begin{equation} \label{useful_result_step}
 \left[a_\alpha,\hat{\mathcal{Y}}_{lm}\right]=l\delta_{\alpha 1}c_{lm}\hat{L}_-^{l-m}\left(a_1^\dagger\right)^{l-1}\left(a_2\right)^l+(l-m)\hbar\delta_{\alpha 2}\frac{c_{lm}}{c_{l,m+1}}\left[a_1,\hat{\mathcal{Y}}_{l,m+1}\right]
\end{equation}
For $\alpha=1$ and $m\rightarrow m+1$
\begin{equation}
 \left[a_1,\hat{\mathcal{Y}}_{l,m+1}\right]=lc_{l,m+1}\hat{L}_-^{l-m-1}\left(a_1^\dagger\right)^{l-1}\left(a_2\right)^l
\end{equation}
Putting this back in (\ref{useful_result_step}) gives
\begin{equation}
 \left[a_\alpha,\hat{\mathcal{Y}}_{lm}\right]=lc_{lm}\left(\delta_{\alpha 1}\hat{L}_-^{l-m}\left(a_1^\dagger\right)^{l-1}\left(a_2\right)^l+(l-m)\hbar \delta_{\alpha 2}\hat{L}_-^{l-m-1}\left(a_1^\dagger\right)^{l-1}\left(a_2\right)^l\right)
\end{equation}
Multiplying (\ref{a_alpha_dagger_fr_com}) on the left gives
\begin{eqnarray}
 &\left[a_\alpha^\dagger, \hat{g}(\hat{R})\right]\left[a_\alpha,\hat{\mathcal{Y}}_{lm}\right]& \nonumber \\
&=& \nonumber \\
&lc_{lm}\left(\hat{g}\left(\hat{R}-\frac{\theta}{2}\right)-\hat{g}(\hat{R})\right)\left(a_1^{\dagger}\hat{L}_-^{l-m}\left(a_1^\dagger\right)^{l-1}\left(a_2\right)^l+(l-m)\hbar a_2^{\dagger}\hat{L}_-^{l-m-1}\left(a_1^\dagger\right)^{l-1}\left(a_2\right)^l\right)& \label{appendix_calculation}
\end{eqnarray}
We note
\begin{eqnarray}
 \left[a_1^\dagger,\hat{L}_-\right]=-\hbar a_2^\dagger, \quad \quad
  \left[a_2^\dagger,\hat{L}_-\right]=0 &\quad \quad \Rightarrow \quad \quad &
   \left[a_1^\dagger,\hat{L}_-^n\right]=-n\hbar a_2^\dagger\hat{L}_-^{n-1}
\end{eqnarray}
In particular for $n=l-m$
\begin{equation}
a_1^\dagger\hat{L}_-^{l-m}=\hat{L}_-^{l-m}a_1^\dagger-(l-m)\hbar a_2^\dagger\hat{L}_-^{l-m-1}
\end{equation}
We put this in equation (\ref{appendix_calculation}) to get
\begin{equation} \label{adagger_f_a_y}
 \left[a_\alpha^\dagger,\hat{g}(\hat{R})\right]\left[a_\alpha,\hat{\mathcal{Y}}_{lm}\right]=l\left(\hat{g}\left(\hat{R}-\frac{\theta}{2}\right)-\hat{g}(\hat{R})\right)\hat{\mathcal{Y}}_{lm}
\end{equation}
In an essentially similar manner one can also prove
\begin{equation} \label{a_f_adagger_y}
 \left[a_\alpha,\hat{g}(\hat{R})\right]\left[a_\alpha^\dagger,\hat{\mathcal{Y}}_{lm}\right]=l\left(\hat{g}(\hat{R})-\hat{g}\left(\hat{R}+\frac{\theta}{2}\right)\right)\hat{\mathcal{Y}}_{lm}
\end{equation}

\section{Properties of the mixed spherical harmonics $\hat{\mathcal{Y}}_{lm}$} \label{app_mixed_Ylm}
\label{A}
From (\ref{l-}) we see
\begin{equation}
\hat{L}_- = \hat{A} + \hat{B}
\end{equation}
with
\begin{equation}
\hat{A} \hat{\psi} = \hbar a_2^{\dagger}[a_1,\hat{\psi}], \quad \quad \hat{B}\hat{\psi} = \hbar [a_2^{\dagger},\hat{\psi}]a_1
\end{equation}
One can check that
\begin{equation}
[\hat{A}, \hat{B}] = 0
\end{equation}
Hence we get
\begin{equation}
\hat{L}_-^n = \left(\hat{A} + \hat{B}\right)^n = \sum_{s=0}^{n}
\left(
\begin{array}{c}
n \\
s
\end{array}
\right)
\hat{A}^s \hat{B}^{n-s}
\end{equation}
Now it is easy to show (try mathematical induction)
\begin{equation}
\hat{B}^n \hat{\mathcal{Y}}_{ll} = (-1)^n l(l-1)...(l-n+1)\hbar^n \left(a_1^\dagger\right)^l a_2^{l-n} a_1^n
\end{equation}
and further
\begin{equation}
\hat{A}^m \hat{B}^n \hat{\mathcal{Y}}_{ll} = (-1)^n \,\, l(l-1)...(l-n+1) \quad l(l-1)...(l-m+1) \hbar^{n+m} \left(a_2^\dagger\right)^m \left(a_1^\dagger\right)^{l-m} a_2^{l-n} a_1^n
\end{equation}
Thus we get
\begin{eqnarray}
\hat{\mathcal{Y}}_{lm} &=& c_{lm} \hat{L}_-^{l-m} \hat{\mathcal{Y}}_{ll} \nonumber \\
&=& c_{lm} \sum_{s=0}^{l-m}
\left(
\begin{array}{c}
l-m \\
s
\end{array}
\right)
(-1)^{l-m-s} \,\, l(l-1)...(m+s+1) \nonumber \\ 
&& \hspace{1.5 cm} l(l-1)...(l-s+1) \hbar^{l-m} \left(a_2^\dagger\right)^s \left(a_1^\dagger\right)^{l-s} a_2^{m+s} a_1^{l-m-s} \nonumber \\
&=& c_{lm} \sum_{s=max(0,-m)}^{min(l-m,l)} (-1)^{l-m-s} \frac{(l-m)! \,\, l! \,\, l!}{s! (l-m-s)! (m+s)! (l-s)!}
\hbar^{l-m} \left(a_2^\dagger\right)^s \left(a_1^\dagger\right)^{l-s} a_2^{m+s} a_1^{l-m-s} \nonumber \\ 
&=& (-\hbar)^{l-m} (l-m)! \,\, c_{lm} \sum_{s=max(0,-m)}^{min(l-m,l)} (-1)^{s} 
\left(
\begin{array}{c}
l \\
m+s
\end{array}
\right)
\left(
\begin{array}{c}
l \\
s
\end{array}
\right)
\left(a_2^\dagger\right)^s \left(a_1^\dagger\right)^{l-s} a_2^{m+s} a_1^{l-m-s} \nonumber \\ 
\end{eqnarray}
It's easy to see that
\begin{equation} \label{ylm_kernel}
\hat{\mathcal{Y}}_{lm} |n_1,n_2\rangle = 0 \quad \quad {\rm for} \,\, n=n_1+n_2<l
\end{equation}
Also note that $\hat{\mathcal{Y}}_{lm}$ does not change the value of $n_1+n_2$ when acting on $|n_1,n_2\rangle$.


\begin{thebibliography}{999}

\bibitem{Doplicher} S.Doplicher, K.Fredenhagen and J.E.Roberts, Comm.Math.Phys. 172,187 (1995).

\bibitem{scholtz1} F.G.Scholtz, L.Gouba, A.Hafver, C.M.Rohwer J.Phys.A 42,175303 (2009).

\bibitem{Balachandran:2004rq} 
  A.~P.~Balachandran, T.~R.~Govindarajan, C.~Molina and P.~Teotonio-Sobrinho,
  JHEP {\bf 0410}, 072 (2004)
  [hep-th/0406125].

\bibitem{doug} M.R. Douglas and N.A. Nekrasov, Rev. Mod. Phys. {\bf 73} (2001) 977.

\bibitem{scholtz2} F.G. Scholtz, B. Chakraborty, J. Govaerts and S. Vaidya, J. Phys. A {\bf 40} (2007) 14581.

\bibitem{scholtz3}  F.G. Scholtz and J Govaerts, Jnl. Phys. A {\bf 41} (2008) 505003.

\bibitem{scholtz4} J.N. Kriel and F.G. Scholtz 2012, Jnl. Phys. A {\bf 45} (2012) 095301. 

\bibitem{press} V. G\'alikov\'a and P. Presnajder,  arXiv:1302.4623.

\bibitem{scholtz5} H.W. Groenewald, J.N. Kriel and F.G.Scholtz, in preparation.

\bibitem{krall} H. L. Krall and Orrin Frink, Transactions of the American Mathematical Society, Vol. 65, No. 1 (Jan., 1949), pp. 100.

\bibitem{Jagerman} D. L. Jagerman, {\em Difference equations with applications to queues},\\Marcel Dekker Inc., (2000). See chapter 7.

 \bibitem{Galikova:2011zf}  V.~Galikova and P.~Presnajder, arXiv:1112.4643 [math-ph].
 
   \bibitem{abromowitz} M. Abromowitz and I. A. Stegun, {\em Handbook of Mathematical Functions},\\Dover Publications, (1972).

\bibitem{Gregg:2008jb} 
  M.~Gregg and S.~A.~Major,
  Int.\ J.\ Mod.\ Phys.\ D {\bf 18}, 971 (2009)
  [arXiv:0806.3496 [astro-ph]].
  
\bibitem{Camacho:2007qy} 
  A.~Camacho and A.~Macias,
  Gen.\ Rel.\ Grav.\  {\bf 39}, 1175 (2007)
  [gr-qc/0702150 [GR-QC]].
  
\bibitem{Chandra:2011nj} 
  N.~Chandra and S.~Chatterjee,
  Phys.\ Rev.\ D {\bf 85}, 045012 (2012)
  [arXiv:1108.0896 [gr-qc]].
  
  	\bibitem{szego} G. Szeg\"{o}, {\em Orthogonal polynomials},\\ American Mathematical Soc., (1939).
  
\bibitem{AmelinoCamelia:2000mn}
  G.~Amelino-Camelia,
  Int.\ J.\ Mod.\ Phys.\  {\bf D11}, 35-60 (2002).
  [gr-qc/0012051].
  
\bibitem{Magueijo:2002am}
  J.~Magueijo, L.~Smolin,
  Phys.\ Rev.\  {\bf D67}, 044017 (2003).
  [gr-qc/0207085].
  
\bibitem{KowalskiGlikman:2001ct}
  J.~Kowalski-Glikman,
  Phys.\ Lett.\  {\bf A299}, 454-460 (2002).
  [hep-th/0111110].
  
%
%
%
%
%
	
\end{thebibliography}
\end{document}